\documentclass[aps, pra, twocolumn, showpacs,superscriptaddress,10pt]{revtex4-1}

\usepackage{amsfonts,amssymb,amsmath}
\usepackage{graphics,graphicx,epsfig}
\usepackage{amsthm}
\usepackage{color}
\usepackage{epstopdf}

\def\identity{\leavevmode\hbox{\small1\kern-3.8pt\normalsize1}}

\newtheorem{lemma}{Lemma}
\newtheorem{remark}{Remark}
\newtheorem{propo}{Proposition}
\newcommand{\be}{\begin{eqnarray}}
\newcommand{\ee}{\end{eqnarray}}
\newcommand{\bpr}{\begin{propo}}
\newcommand{\epr}{\end{propo}}
\newcommand{\ble}{\begin{lemma}}
\newcommand{\ele}{\end{lemma}}
\newcommand{\bpf}{\begin{proof}}
\newcommand{\epf}{\end{proof}}
\newcommand{\ket}[1]{\left | #1 \right\rangle}
\newcommand{\bra}[1]{\left \langle #1 \right |}

\newcommand{\Tr}{\mathrm{Tr}}
\newcommand{\Ham}{\mathcal{H}}

\newcommand{\QFI}{\mathcal{F}_Q}
\newcommand{\diff}{\operatorname{d}}

\newcommand{\proj}[1]{\ket{#1}\bra{#1}}
\renewcommand{\epsilon}{\varepsilon}

\def\openone{\leavevmode\hbox{\small1 \normalsize \kern-.64em1}}

\usepackage[T1]{fontenc}

\begin{document}

\title{Quantum noise generated by local random Hamiltonians}

\author{Marcin Markiewicz} \email{marcin.1.markiewicz@uj.edu.pl}    
\affiliation{Institute of Physics, Jagiellonian University, \L{}ojasiewicza 11,
30-348 Krak\'ow, Poland} 

\author{Zbigniew Pucha{\l}a}
\affiliation{Institute  of  Theoretical  and  Applied  Informatics,
Polish  Academy  of  Sciences,  Ba{\l}tycka  5,  44-100  Gliwice,  Poland}
\affiliation{Institute of Physics, Jagiellonian University, \L{}ojasiewicza 11,
30-348 Krak\'ow, Poland} 

\author{Anna de Rosier}
\affiliation{Institute of Theoretical Physics and Astrophysics, University of Gda\'nsk, 80-952 Gda\'nsk, Poland}

\author{Wies{\l}aw Laskowski}
\affiliation{Institute of Theoretical Physics and Astrophysics, University of Gda\'nsk, 80-952 Gda\'nsk, Poland}

\author{Karol \.Zyczkowski}     
\affiliation{Institute of Physics, Jagiellonian University, \L{}ojasiewicza 11,
30-348 Krak\'ow, Poland} 
\affiliation{Center for Theoretical Physics, Polish Academy of Sciences, Warsaw, Poland}

\date{\today}


\begin{abstract}
We investigate the impact of a local random unitary noise on multipartite quantum states of an arbitrary dimension. We follow the dynamical approach, in which the single-particle unitaries are generated by local random Hamiltonians. Assuming short evolution time we derive a lower bound on the fidelity between an initial and the final state transformed by this type of noise. This result is based on averaging the Tamm-Mandelstam bound and holds for a wide class of distributions of random Hamiltonians fulfilling specific symmetry conditions. 
It is showed how the sensitivity of a given pure quantum state to the discussed type of noise depends on the properties of single-particle and bipartite reduced states.
\end{abstract}


\maketitle

\section{Introduction}

Fiducial transmission of quantum particles is a crucial element of many quantum information protocols \cite{Gisin07}. In this work we discuss the case of transmitting an arbitrary number $n$ of $d$-level particles which undergo a local unitary noise. This kind of noise is typical for optical implementations of many quantum protocols \cite{Wang05}. We compare two scenarios. First, in which the particles are send in a single wave-packet (each particle undergoes the same noise), and the second one, in which they are send sequentially (each particle undergoes a noise sampled independently from the same distribution). We introduce a characterization of quantum states with respect to a sensitivity to such a noise by means of the so called Quantum Fisher Information (QFI).

Random unitary noise of the general form:
\be
\mathbb{V}[\rho]=\int \mathcal{V}\rho\mathcal{V}^{\dagger}\diff \mathcal V,
\label{GenNoise}
\ee
where $\mathcal V$ is some unitary matrix, can acquire different forms, with respect to two aspects: locality and the way of sampling the unitary operators. The locality aspect tells us whether the noise acts independently on the parties: $\mathcal V=U_1\otimes\ldots\otimes U_n$ with $U_i\in\operatorname{SU}(d)$, or it has the global character: $\mathcal V\in \operatorname{SU}(d^n)$. When it comes to the second aspect, one can sample the unitary from the corresponding unitary group according to some convenient measure (so-called \emph{twirling}), or one can adapt the dynamical approach and assume that a random unitary is generated by a random Hamiltonian: $\mathcal V=e^{-it\Ham}$. In the latter case the quantum channel \eqref{GenNoise} is actually a continuous family of channels parametrized by the interaction time $t$. To sum up, we distinguish four main types of random unitary noise: (i) global twirling, (ii) local twirling, (iii) global random Hamiltonian model and (iv) local random Hamiltonian model. 

\emph{Global twirling} channel is actually equivalent to completely depolarizing channel \cite{Emerson05, Bartlett07}. Global noise generated by a random Hamiltonian, $\mathcal V=e^{-itH}$, gives rise to a partial depolarization of the $n$-partite state $\rho(t)=p(t)\rho_0+(1-p(t))\openone$, if one assumes that the distribution of eigenvectors of $H$ is invariant under unitary conjugation \cite{RandHam12}. This type of evolution was studied in the theory of open quantum systems in the context of a reduced random dynamics \cite{Znidaric2011, Brandao2012, Gessner2013}.
\emph{Local twirling} \cite{Bartlett07} has been mainly considered in the case of identical local unitaries: $\mathcal V=U^{\otimes n}$. In this case the form of the output state can be derived from the multiplet structure of $n$ composed spins-$j$, in which $j=(d-1)/2$. This model describes a noise which appears due to the lack of a shared reference frame between sender and receiver in quantum communication \cite{Bartlett03, Gour09, Bartlett14}. The states which are invariant with respect to a local twirling are called \emph{isotropic states} \cite{Werner89, HH99, Johnson13}.

Our paper is focused on the fourth possibility: \emph{local noise generated by a local random Hamiltonian}, $\mathcal V=e^{-it\mathcal H}$, where the Hamiltonian $\Ham$ has a direct sum structure. We develop an efficient method for describing the impact of this type of noise on  quantum states.
We describe the effect of the random noise using the fidelity between the initial state and the state at the final  time $t$.
Direct calculation of the fidelity can be intractable for a large number $n$ of particles  and higher number $d$ of levels, especially in the case of mixed states. This difficulty can be overcome by bounding the fidelity with Quantum Fisher Information (QFI) for sufficiently small evolution time intervals, in which the fidelity is an invertible function of time.
QFI, originally developed as a mean to describe the efficiency of a parameter estimation in quantum metrology \cite{Helstrom67, Braunstein94, FisherDef}, was further applied to quantify performance of some information processing tasks, like imperfect quantum cloning \cite{clone1, clone2}, adjusting reference frames \cite{Ahmadi15} and quantum search \cite{ourGrover}. An analysis of QFI for random quantum pure states was recently presented in \cite{Oszmaniec2016}. 

In this work we apply QFI to the problem of bounding the fidelity of a multipartite quantum state undergoing random unitary noise. The key idea is that optimizing QFI over different directions corresponding to generators of a unitary group in the case of qubits is very simple. We generalize this idea to the case of arbitrary finite-dimensional systems, and show, how this property of QFI allows one to assess the fidelity. 

\section{Basic definitions}

\subsection{Random local Hamiltonians}

In this work we consider a system of $n$ particles with $d$-levels each, undergoing two types of a random unitary noise generated by random local  Hamiltonians. In the first case we assume that the Hamiltonian is collective, which means that each particle is affected by the same noise:
\be
\Ham_{\vec k} = H_{\vec k}\otimes \openone^{\otimes (n-1)}+\openone\otimes H_{\vec k}\otimes\openone^{\otimes (n-2)}+\ldots, 
\label{genHcol}
\ee
(we will use ordinary capital letters for single-particle operators, and \emph{mathcal} letters for $n$-particle operators). The subscript $\vec k$ denotes some way of sampling the Hamiltonian, which will be specified later. 
The collective noise appears in different physical scenarios, for example in the case when an external interaction is localised in space and  the particles are send in a single narrow wave-packet. Another example is the situation in which the particles are truly indistinguishable and therefore cannot be individually addressed.
In the second case we assume that the Hamiltonian also has a direct sum structure, however it is sampled independently for each particle. This method of sampling physically corresponds to sending distinguishable particles one by one in a sequence. In this case each subsystem undergoes the  noise by the same amount of time, but independently from the other ones:
\be
\Ham_{\vec k_1,\ldots,\vec k_n} = H_{\vec k_1}\otimes \openone^{\otimes (n-1)}+\openone\otimes H_{\vec k_2}\otimes\openone^{\otimes (n-2)}+\ldots.\nonumber\\ 
\label{genHncol}
\ee
In the above formula different subscripts $\vec k_1,\ldots,\vec k_n$ correspond to the fact that the generator of the noise is sampled independently for each particle, but from the same distribution.

Let us assume that each local Hamiltonian $H_{\vec k}$ belongs to an 
$r$-dimensional subspace of the space of Hermitian operators, and therefore can be expanded 
in some orthogonal set consisting of $r\leq d^2$ local Hamiltonians 
$\{H_i\}_{i=1}^r$ (for example Pauli matrices, Gell-Mann matrices, spin 
matrices):
\be
H_{\vec k}=\sum_{i=1}^r \alpha_i(\vec k) H_i,
\label{localHam}
\ee
for which we assume that $Tr(H_i^2)=\textrm{const}$ for each index $i$. The set of real coefficients $\vec\alpha_{\vec k}\equiv[\alpha_1(\vec k),\ldots,\alpha_r(\vec k)]$ representing the Hamiltonian $H_{\vec k}$ can be treated as a real $r$-dimensional vector. We assume that the distribution of the Hamiltonian is represented by a measure $\diff \vec k$, invariant with respect to orthogonal transformations on its vectorized representation $\vec\alpha_{\vec k}$:
\be
\int\vec\alpha_{\vec k}\diff\vec k=\int\hat O\vec\alpha_{\vec k}\diff\vec k,
\label{HamSym}
\ee
where $\hat O$ is any element from the $r$-dimensional orthogonal group 
$\textrm{O}(r)$. The above construction is a slight generalization of the 
typical random matrix ensambles, like Gaussian Unitary Ensamble (GUE) and  
Gaussian Orthogonal Ensamble (GOE). If we assume, that the vector $\vec \alpha$ is a
multivariate standard normal then we obtain the distribution of GUE restricted to a subspace 
spanned by $H_i$, see discussion in Appendix~\ref{sec:generalizedGUE}. If we additionally assume that the set of operators $\{H_i\}_{i=1}^r$ consists of real symmetric matrices, then we obtain the distribution of GOE.
Several distributions are discussed in details in section \ref{SecDist}.

\subsection{Collective and non-collective channels generated by random local Hamiltonians}

The Hamiltonians \eqref{genHcol} and \eqref{genHncol} generate two families of quantum channels, parametrized by the time $t$,
\be
\mathcal U^{\textrm{c}}_{t}[\rho]=\int \diff\vec k\,\, e^{-i t \Ham_{\vec k}}\rho \,\,e^{i t \Ham_{\vec k}},
\label{genUS2}
\ee
\be
\mathcal U^{\textrm{nc}}_{t}[\rho]=\int \diff\vec k_1\ldots\diff\vec k_n\,\, e^{-i t \Ham_{\vec k_1\ldots\vec k_n}}\rho \,\,e^{i t \Ham_{\vec k_1\ldots\vec k_n}}.
\label{genUSd}
\ee
Although channel \eqref{genUS2} seems to be very similar to the twirling channel of the general form:
\be
\mathcal U^{\textrm{twirl}}[\rho]=\int \diff U \,\, U^{\otimes n}\rho \,\,U^{\otimes n},
\label{twirlgen}
\ee
both maps have different properties. Firstly, the twirling channel is a projector:
\be
\mathcal U^{\textrm{twirl}}(\mathcal U^{\textrm{twirl}}[\rho])=\mathcal U^{\textrm{twirl}}[\rho],
\label{twirl1}
\ee
and its direct form can be found from the multiplet structure of the $n$ composed spins $(d-1)/2$. For example in the case of $d=2$ the twirling channel has a direct closed form \cite{Bartlett03}:
\be
\mathcal U^{\textrm{twirl}}=\sum_{j=0}^{n/2}(\mathcal D_{\mathcal M_j}\otimes\openone_{\mathcal N_j})\circ\mathcal P_j,
\label{twirl2}
\ee
where $\mathcal D_{\mathcal M_j}$ is a depolarization channel on the so called \emph{gauge subspace}, on which an irreducible representation of a unitary group $\textrm{SU}(2)$ acts, $\mathcal N_j$ is a \emph{multiplicity subspace} on which the trivial representation of $\textrm{SU}(2)$ acts, and $\mathcal P_j$ is a projector onto an invariant subspace corresponding to spin-$j$.
In contrast the operation \eqref{genUS2} is neither a projector, nor it forms a dynamical semigroup. However, it respects the invariant subspace structure of the twirling channel \eqref{twirl2}, and in some cases it gives rise to a periodic evolution. 

Let us discuss this aspect more thoroughly. As a pedagogical example we take the case of a five-qubit system, for which the six-dimensional fully symmetric subspace is spanned by two five-qubit GHZ-type states:
\be
\ket{GHZ^{\pm}_5}=\frac{1}{\sqrt{2}}(\ket{00000}\pm\ket{11111}),
\label{GHZdef}
\ee
and by four Dicke states:
\begin{equation}
\ket{D^e_5} = \frac{1}{\sqrt{{5 \choose e}}} \sum_{\pi} \ket{\pi(1 \dots 1\,\, 0 \dots 0)},\,\,\,e=1,2,3,4,
\label{DEN}
\end{equation}
where $\ket{0}$ and $\ket{1}$ are the eigenstates of the local Pauli operator $\sigma_z$ corresponding to  eigenvalues $\pm 1$. Here 
$\pi$ denotes a permutation of $e$ ones and $(5-e)$ zeros in the ket.
Assume we take $\ket{GHZ^+_5}$  as the initial state, and we sample the Hamiltonian from the uniform distribution on a two-dimensional sphere: 
$H_{\vec k}=\sum_{i=1}^3 k_i \sigma_i$, in which  $\vec k\in S^2$ and $\sigma_i$ denote Pauli matrices.
After the action of the transformation \eqref{genUS2} the state $\ket{GHZ^+_5}$ is transformed into a mixture of all the basis vectors of the symmetric subspace:
\be
&&\ket{GHZ^+_5}\stackrel{\mathcal U_{t}^{\textrm{c}}}{\longrightarrow}\zeta_1(t)\left(\proj{D^1_5}+\proj{D^4_5}\right)+\nonumber\\
&&\zeta_2(t)\left(\proj{D^2_5}+\proj{D^3_5}\right)+\zeta_3(t)\proj{GHZ^+_5}\nonumber\\
&&+\zeta_4(t)\proj{GHZ^-_5},
\label{evolvedGHZ}
\ee
in which all the probabilities $\zeta_i(t)$ are periodic functions of time -- see Appendix \ref{GHZ5App}. On the other hand, the twirling channel \eqref{twirlgen} transforms the GHZ state to a projector onto the fully symmetric subspace, which corresponds to all $\zeta_i(t)=\frac{1}{6}$ in \eqref{evolvedGHZ}. As shown in  Appendix \ref{GHZ5App}, the coefficients $\zeta_i(t)$ in \eqref{evolvedGHZ} are never equal, therefore the operation \eqref{genUS2}  does not reproduce the output of the twirling channel \eqref{twirlgen} for any time instant. 

\subsection{Fidelity and Quantum Fisher Information}

As a measure of the quality of a state transfer through a channel parametrized by time, we take the Bures fidelity between input and output states \cite{Bures69, Helstrom67, Jozsa94, Bengtsson2006}:
\be
&&F_B(\rho,\rho_t)=F_B(\rho,\mathcal U_{t}[\rho])=\left(\Tr \left(\sqrt{\sqrt{\rho}\cdot\rho_{t}\cdot\sqrt{\rho}}\right)\right)^2.\nonumber\\
\label{BuresFid}
\ee
In order to find analytically a tractable bound on fidelity for random unitary noise generated by random Hamiltonians we utilize the relation between fidelity and Quantum Fisher Information (QFI) \cite{Braunstein94, Taddei13}. Indeed, let us assume that the state $\rho_{0}$ undergoes a unitary evolution $U_t=e^{-it\Ham}$ generated by a Hamiltonian $\Ham$.
The fidelity between the initial and evolved state can be expanded in a power series with respect to $t$ \cite{Taddei13}:
\be
F_B(\rho,\rho_t)=1-\QFI(\rho,\mathcal H)\frac{t^2}{4}+\operatorname{O}(t^3),
\label{PSQFI}
\ee
in which the coefficient in front of $t^2$ is given by the Quantum Fisher Information $\QFI$.
For a general input state $\rho$ with the eigendecomposition $\rho=\sum_i \lambda_i |i\rangle\langle i|$ the QFI takes the following form \cite{Helstrom67, Braunstein94}:
\be
\QFI(\rho,\Ham)=2\sum_{m,l}\frac{(\lambda_m-\lambda_{l})^2}{\lambda_m+\lambda_{l}}|\langle m|\Ham|l\rangle |^2,
\label{qfidef}
\ee
in which the sum runs over eigenvalues for which the denominator is non-zero. Note that the above formula implies that QFI is invariant with respect to shifting the unitary dynamics from the state to the Hamiltonian:
\be
\QFI(U\rho U^{\dagger},\Ham)=\QFI(\rho,U^{\dagger}\Ham U),
\label{QFIpictures}
\ee
which resembles the equivalence between Schroedinger and Heisenberg pictures in quantum mechanics.
It can be seen that the QFI describes the speed of divergence of quantum states $\rho$ and $\rho_t$ under the unitary evolution, $U_t=e^{it\Ham}$. 
Apart from \eqref{PSQFI} a direct bound on fidelity using the QFI can be found \cite{Mandelstam45, Frowis12}:

\be
F_B(\rho,\rho_t)\geq \cos^2\left(\frac{t}{2}\sqrt{\QFI(\rho,\Ham)}\right),
\label{qfifid}
\ee 
which holds for small time intervals \cite{Frowis12}:
\be
|t|\leq t^*=\frac{\pi}{\sqrt{\QFI(\rho,\Ham)}}.
\label{qfifidas}
\ee
The above condition implies that the evolution time is sufficiently short to assure the uniqueness of the fidelity as a function of time -- for longer evolution times the evolving state can return closer to the initial point. The bound \eqref{qfifid} is known in the literature as the Tamm-Mandelstam bound, since it has been already proposed for pure states in the fourties \cite{Mandelstam45}, a long time before the Quantum Fisher Information was defined.

It is worth mentioning that the bound \eqref{qfifid}, when put into a bit different form:
\be  
t\geq \frac{\arccos\left(\sqrt{F_B(\rho,\rho_t)}\right)}{\sqrt{\frac{1}{4}\QFI(\rho,\Ham)}},
\label{QSL1}
\ee
is an example of a \emph{Quantum Speed Limit} (QSL) \cite{Giovannetti03}, and can be interpreted as a lower bound on the time in which a given initial state $
\rho$ can evolve to a final state $\rho_t$ under unitary evolution generated by a Hamiltonian $\Ham$. A more general form of QSL has been recently introduced in \cite{Pires16}:
\be
t\geq \frac{\mathcal L(\rho,\rho_t)}{\mathcal S(\rho,\rho_t)},
\label{QSLgen}
\ee
in which $\mathcal S(\rho,\rho_t)$ is a length of the path in the operator space followed by a state inbetween $\rho$ and $\rho_t$, calculated according to a specific metric tensor $g_{\mu\nu}$  (see eq. (21) in \cite{Pires16}), whereas $\mathcal L(\rho,\rho_t)$ is a geodesic distance between $\rho$ and $\rho_t$ corresponding to this metric. In the case of QSL in the form \eqref{QSL1}  the length $\mathcal S(\rho,\rho_t)$ is calculated according to a Fisher Information metric:
\be
[\QFI(\rho)]_{\mu,\nu}=\frac{1}{2}\sum_{m,l}\frac{(\lambda_m-\lambda_{l})^2}{\lambda_m+\lambda_{l}}\langle m|\Ham_{\mu}|l\rangle\langle l|\Ham_{\nu}|m\rangle,\nonumber\\
\label{qfimetric}
\ee
whereas the geodesic distance $\mathcal L(\rho,\rho_t)$ corresponds to the Bures angle \cite{Bures69, Bengtsson2006}:
\be  
\mathcal L(\rho,\rho_t)=\arccos\left(\sqrt{F_B(\rho,\rho_t)}\right).
\label{BuresAng}
\ee
In principle one can take any of the infinite family of speed limits \eqref{QSLgen}  and try to translate it into a form analogous to \eqref{qfifid}. However in two special cases  the form of a geodesic distance $\mathcal L$ is known analytically: for the Bures angle \eqref{BuresAng}, connected with Bures fidelity $F_B$ \eqref{BuresFid}, and for the Wigner-Yanase distance \cite{WYdist}:
\be  
\mathcal L_{WY}(\rho, \rho_t)=\arccos(\Tr(\sqrt{\rho}\sqrt{\rho_t})).
\label{WYdist}
\ee
The corresponding measure of distinguishability, analogous to Bures fidelity, is in the latter case given by the  \emph{quantum affinity}  \cite{Afidelity}:
\be  
A(\rho, \rho_t)=\Tr(\sqrt{\rho}\sqrt{\rho_t}).
\label{Afid}
\ee
It allows one to derive the following quantum speed limit \cite{Pires16}:
\be
t\geq \frac{\arccos(A(\rho, \rho_t))}{\frac{\sqrt{2}}{t}\int_{0}^t\diff\tau\sqrt{I(\rho_{\tau}, \Ham)}},
\label{QSLWY}
\ee
in which the quantity: 
\be
I(\rho_{\tau},\Ham)=-\Tr\left((\sqrt{\rho_{\tau}}\Ham-\Ham\sqrt{\rho_{\tau}})^2\right), 
\label{Skew}
\ee
analogous  to the Quantum Fisher Information, is known as the Quantum Skew Information \cite{WYdist}. The corresponding bound for the quantum affinity \eqref{Afid} reads:
\be
A(\rho, \rho_t)\geq \cos\left(\sqrt{2}\int_0^t \diff \tau \sqrt{I(\rho_{\tau}, \Ham)}\right),
\ee
and is  more difficult to calculate than  the Tamm-Mandelstam bound \eqref{qfifid}.

\section{Main results}

Our main result relies on providing an efficient method of averaging the Tamm-Mandelstam bound \eqref{qfifid}. Namely we will show in section \ref{sec:bound}, that if instead of a unitary evolution generated by a single Hamiltonian $\Ham$ we take noisy channels  \eqref{genUS2} and \eqref{genUSd}, the bound \eqref{qfifid} is still valid:
\be
F_B\left(\rho, \, \int e^{-it\Ham_{\vec k}}\rho e^{it\Ham_{\vec k}}\diff \vec k\right)\geq \cos^2\left(\Omega t\right)
\label{MainBoundv1}
\ee
on condition that we also average the QFI:
\be
\Omega=\frac{1}{2}\sqrt{\int\QFI(\rho,\Ham_{\vec k})\diff\vec k}.
\label{meanQFI}
\ee
The quantity $\Omega$, which plays a role of a frequency of the evolution of quantum states under the action of \eqref{genUS2} and \eqref{genUSd}, is determined by the mean QFI averaged over the space of Hamiltonians \eqref{localHam} according to a measure $\diff \vec k$, which fulfills the symmetry conditions \eqref{HamSym}. The averaged bound \eqref{MainBoundv1} holds for small evolution times fulfilling $t\leq\pi/(2\Omega)$ in full analogy to \eqref{qfifidas}.

Therefore we translate the problem of bounding the fidelity of channels \eqref{genUS2} and \eqref{genUSd}  into the problem of averaging the QFI. We show in sections \ref{sec:col} and \ref{sec:noncol} that the  quantity \eqref{meanQFI} can be easily calculated for any finite dimensional quantum system and any method of sampling the Hamiltonian fulfilling the symmetry conditions \eqref{HamSym}. In the case of sampling the local  Hamiltonian in a collective way \eqref{genUS2}, the mean QFI \eqref{meanQFI} is proportional to the arithmetic mean of the QFI with respect to a local Hermitian basis (details in Proposition \ref{propc}):
\be
\int \QFI(\rho,\Ham_{\vec k})\diff \vec k\,\, \propto\,\,\frac{1}{r}\sum_{i=1}^r \QFI(\rho,\Ham_i).\nonumber\\
\label{meanQFIcol}
\ee
On the other hand in the case of non-collective way of sampling local Hamiltonians \eqref{genUSd}, the mean \eqref{meanQFI} is proportional to the arithmetic mean of the QFI, however calculated in a different way (details in Proposition \ref{propnc}):
\be
\int \QFI(\rho,\Ham_{\vec k_1,\ldots,\vec k_n})\diff \vec k_1\ldots\diff \vec k_n\propto\frac{1}{r}\sum_{\pi}\sum_{i=1}^r\QFI(\rho,\Ham_{\pi(i)}),\nonumber\\
\label{meanQFIncol}
\ee
where:
\be
\Ham_{\pi(i)}=H_{i_1}\otimes H_{i_2}\otimes\ldots\otimes H_{i_n},
\ee
for $\{i_s\}_{s=1}^n=\pi(1,0,\ldots,0)$. In the simplest case of two qubits, for which we take the Pauli matrices $\{\sigma_x,\sigma_y,\sigma_z,\openone\}$ as the local basis, the mean \eqref{meanQFIcol} reads:
\be
\frac{1}{r}\sum_{i=1}^r \QFI(\rho,\Ham_i)=\frac{1}{3}\sum_{i=1}^3\QFI(\rho,\sigma_i\otimes\openone+\openone\otimes\sigma_i).\nonumber\\
\ee
The second average in Eq.  \eqref{meanQFIncol} reads:
\be
\frac{1}{r}\sum_{\pi,i}\QFI(\rho,\Ham_{\pi(i)})&=&\frac{1}{3}\sum_{i=1}^3(\QFI(\rho,\sigma_i\otimes\openone)\nonumber\\
&+&\QFI(\rho,\openone\otimes\sigma_i)).
\ee
The above properties of a Quantum Fisher Information were formerly proven only for the collective Hamiltonian in the case of many qubits \cite{Hyllus10, Toth12}. The equations \eqref{meanQFIcol} and \eqref{meanQFIncol} are analogous in spirit to quantum designs \cite{designs1, designs2}, in which averages over some continuous operator sets are expressed as finite sums. Moreover in section \ref{pures} we show that in the case of pure states the mean QFI's \eqref{meanQFIcol} and \eqref{meanQFIncol} are determined by the local Bloch vectors and bipartite correlations only.

Note that the properties  \eqref{meanQFIcol} and \eqref{meanQFIncol} depend weakly on the way of sampling the Hamiltonians. If the distribution of the local  Hamiltonians is symmetric in the sense of \eqref{HamSym}, the results for different ways of sampling differ only by multiplicative constants of proportionality in \eqref{meanQFIcol} and \eqref{meanQFIncol}, which is discussed in section \ref{SecDist}.

Finally in section  \ref{examples} we discuss the robustness of several families of quantum states against the noise due to both channels \eqref{genUS2} and \eqref{genUSd}. Hence one can optimize the way of transferring the particles. Depending on the state one of the two strategies: sending the parties sequentially or in a single packet brings more advantages.

\section{Bound on fidelity based on Quantum Fisher Information}
\label{sec:bound}
In order to derive a useful bound on the fidelity of the channels (\ref{genUS2}) and (\ref{genUSd}) we first use the concavity of the Bures fidelity \cite{Bengtsson2006}, which implies:
\be 
&F_B\left(\rho, \int \operatorname{d}\vec k\,\, e^{-i t \Ham_{\vec k}}\rho \,\,e^{i t \Ham_{\vec k}}\right)\geq \int\diff \vec k\,\, F_B(\rho, \, e^{-i t \Ham_{\vec k}}\rho \,\,e^{i t \Ham_{\vec k}}).&\nonumber\\
&&
\label{firstapx}
\ee
In the case of a pure input state, inequality  \eqref{firstapx} is saturated: 
\be
&&F_B(\rho,\rho_t)=\Tr(\rho\rho_t)=\Tr\left(\rho\int \operatorname{d} \vec k\,\, e^{-i t \Ham_{\vec k}}\rho \,\,e^{i t \Ham_{\vec k}}\right)\nonumber\\
&&=\int\Tr\left(\rho\,\, e^{-i t \Ham_{\vec k}}\rho \,\,e^{i t \Ham_{\vec k}}\right)\operatorname{d} \vec k=\nonumber\\
&&=\int F_B\left(\rho,\,\, e^{-i t \Ham_{\vec k}}\rho \,\,e^{i t \Ham_{\vec k}}\right)\operatorname{d} \vec k.
\ee
In the next step we utilize the Tamm-Mandelstam bound \eqref{qfifid}:
\be 
&&\int F_B\left(\rho,\,\, e^{-i t \Ham_{\vec k}}\rho \,\,e^{i t \Ham_{\vec k}}\right)\operatorname{d}\vec k\nonumber\\&&\geq \int \cos^2\left(\frac{t}{2}\sqrt{\QFI(\rho,\Ham_{\vec k})}\right)\diff \vec k\nonumber\\
&&\geq\cos^2\left(\frac{t}{2}\sqrt{\int \QFI(\rho,\Ham_{\vec k})\diff \vec k}\right),
\label{mtbound}
\ee
where in the second estimation we use the fact, that the function $\cos^2(\frac{t}{2}\sqrt{\QFI})$ is a convex function of $\QFI$ for all $t$ fulfilling the condition \eqref{qfifidas}. The last formula shows that for any evolution generated by a random Hamiltonian  the fidelity of the evolved state decreases as $\cos^2(\Omega t)$, in which the square root of the mean QFI plays the role of the frequency of quantum evolution:  $\Omega=\frac{1}{2}\sqrt{\int\QFI(\rho,\Ham_{\vec k})\diff \vec k}$. Below we calculate the quantity $\Omega$ for both Hamiltonians \eqref{genHcol} and \eqref{genHncol} and show that the result does not  depend strongly on the way of sampling the Hamiltonian, as long as the symmetry conditions \eqref{HamSym} hold.

\subsection{Mean Quantum Fisher Information for collective random Hamiltonians of the form \eqref{genHcol}}
\label{sec:col}
\bpr
For arbitrary collective Hamiltonian $\Ham_{\vec k}$ of the form  \eqref{genHcol}, in which local Hamiltonians $H_{\vec k}$, given by \eqref{localHam}, are sampled according to the measure $\diff\vec k$ fulfilling conditions \eqref{HamSym}, the mean $\QFI$ is given by:
\be
\int \QFI(\rho,\Ham_{\vec k})\diff \vec k =\left(\frac{\int\Tr(H_{\vec k}^2)\diff\vec k}{\Tr(H_1^2)}\right)\frac{1}{r}\sum_{i=1}^r \QFI(\rho,\Ham_i),\nonumber\\
\ee
in which $\Ham_i=H_i\otimes \openone^{\otimes (n-1)}+\openone\otimes H_i\otimes\openone^{\otimes (n-2)}+\ldots$, and $\Tr(H_i^2)=\textrm{const}$.
\label{propc}
\epr
\bpf
Since the collective Hamiltonian \eqref{genHcol} is a direct sum of local Hamiltonians \eqref{localHam}, we may introduce an $r$-dimensional collective orthonormal basis $\{\hat\gamma_i\}_{i=1}^r$:
\be
\hat \gamma_i &=& c_{\hat\gamma}\left(H_i\otimes \openone^{\otimes (n-1)}+\openone\otimes H_i\otimes\openone^{\otimes (n-2)}+\ldots\right)\nonumber\\
&=&c_{\hat\gamma}\Ham_i,
\label{gengammac}
\ee
in which, due to the orthonormality condition $\Tr(\hat\gamma_i\hat\gamma_j)=\delta_{ij}$, the normalization coefficient is given by:
\be
c_{\hat\gamma}=\frac{1}{\sqrt{nd^{n-1}\Tr(H_1^2)}}.
\label{gammanormc}
\ee
To calculate the integral $\int \QFI(\rho,\Ham_{\vec k})\diff \vec k$, where $\Ham_{\vec k}$ is given by the formula \eqref{genHcol}, we use direct definition of the QFI \eqref{qfidef}:
\be
\int \QFI(\rho,\Ham_{\vec k})\diff \vec k=\int\sum_{m,l}f_{ml}\bra{m}\Ham_{\vec k}\ket{l}\bra{l}\Ham_{\vec k}\ket{m}\diff \vec k\nonumber\\,
\label{qfiint}
\ee
in which $f_{ml}=\frac{2(\lambda_m-\lambda_l)^2}{\lambda_m+\lambda_l}$. 
We expand the collective Hamiltonians in the formula \eqref{qfiint} in the orthonormal basis \eqref{gengammac}:
\be
\int \QFI(\rho,\Ham_{\vec k})\diff \vec k&=&\int\sum_{m,l=1}^r f_{ml}\bra{m}\left(\sum_i\Tr(\Ham_{\vec k}\hat \gamma_i)\hat \gamma_i\right)\ket{l}\nonumber\\
&\times&\bra{l}\left(\sum_j\Tr(\Ham_{\vec k}\hat \gamma_j)\hat \gamma_j\right)\ket{m}\diff \vec k.
\label{firststep}
\ee
After changing the order of summation and changing the notation of the new basis from $\hat\gamma_i$ to $c_{\hat\gamma}\Ham_i$ this expression reads:
\be
\int \QFI(\rho,\Ham_{\vec k})\diff \vec k&=&c_{\hat\gamma}^2\sum_{i,j=1}^r\left(\sum_{m,l} f_{ml}\bra{m}\Ham_i\ket{l}\bra{l}\Ham_j\ket{m}\right)\nonumber\\
&\times&\int \Tr(\Ham_{\vec k}\hat \gamma_i)\Tr(\Ham_{\vec k}\hat \gamma_j) \diff \vec k.
\label{secondstep}
\ee
The above formula can be presented in a compact form:
\be
&&\int \QFI(\rho,\Ham_{\vec k})\diff \vec k=c_{\hat\gamma}^2 \sum_{i,j=1}^r\Gamma_{ij}\int \Tr(\Ham_{\vec k}\hat \gamma_i)\Tr(\Ham_{\vec k}\hat \gamma_j) \diff \vec k,\nonumber\\
\label{qfimatrix}
\ee
in which:
\be
\Gamma_{ij}\equiv\sum_{m,l} f_{ml}\bra{m}\Ham_i\ket{l}\bra{l}\Ham_j\ket{m},
\label{Gmat}
\ee
is the Fisher Information matrix, used in multiparameter estimation \cite{Kolodynski10, FisherDef}. It indicates how a quantum state is sensitive to a unitary evolution generated by Hamiltonians pointing in different directions in the Hilbert space.
It remains to calculate the integral $\int \Tr(\Ham_{\vec k}\hat \gamma_i)\Tr(\Ham_{\vec k}\hat \gamma_j) \diff \vec k$. Firstly we express the collective operators in terms of single-particle ones:
\be
&&\int \Tr(\Ham_{\vec k}\hat \gamma_i)\Tr(\Ham_{\vec k}\hat \gamma_j) \diff \vec k\nonumber\\
&=&\int \Tr(H_{\vec k}H_i\otimes\openone_{n-1}+\ldots+\openone_{n-1}\otimes H_{\vec k}H_i)\nonumber\\
&\times& \Tr(H_{\vec k}H_j\otimes\openone_{n-1}+\ldots+\openone_{d-1}\otimes H_{\vec k}H_j) \diff \vec k,\nonumber\\
&=&c_{\vec\gamma}^2 n^2d^{2(n-1)}\int\Tr(H_{\vec k}H_i)\Tr(H_{\vec k}H_j) \diff \vec k,
\label{singleint}
\ee
in which the last equality follows from the linearity of trace and factorization property of trace with respect to a tensor product.
The integral can be finally calculated using the following lemma, the proof of which we postpone to Appendix \ref{applemma}:
\ble
Assume that the single-particle Hamiltonian is sampled from the subspace of Hermitian matrices of a real dimension $r$, and that the distribution of $H$ fulfills the symmetry conditions \eqref{HamSym}. Then the integral \eqref{singleint} reads:
\be
\int \Tr(H_{\vec k}H_i)\Tr(H_{\vec k}H_j) \diff \vec k= \delta_{ij}\frac{\Tr(H_1^2)}{r}\int\Tr(H_{\vec k}^2)\diff\vec k.\nonumber
\ee
\label{lem1}
\ele
Using the above lemma we obtain:
\be
\int \Tr(\Ham_{\vec k}\hat \gamma_i)\Tr(\Ham_{\vec k}\hat \gamma_j) \diff \vec k&=&c_{\vec\gamma}^2 n^2d^{2(n-1)}\delta_{ij}\frac{\Tr(H_1^2)}{r}\nonumber\\
&\times&\int\Tr(H_{\vec k}^2)\diff\vec k.
\ee
Putting the above result into formula \eqref{qfimatrix}, we obtain the final form of the integral \eqref{qfiint}:
\be
&&\int \QFI(\rho,\Ham_{\vec k})\diff \vec k = c_{\vec\gamma}^4 n^2d^{2(n-1)}\Tr(H_1^2)\frac{\int\Tr(H_{\vec k}^2)\diff\vec k}{r}\nonumber\\
&&\times\sum_i \Gamma_{ii}\nonumber=\left(\frac{\int\Tr(H_{\vec k}^2)\diff\vec k}{\Tr(H_1^2)}\right)\frac{1}{r}\sum_{i=1}^r \QFI(\rho,\Ham_i),\\
\label{qfifin}
\ee
where in the last step we used the definition \eqref{gammanormc}  of $c_{\vec\gamma}$ and the fact, that the diagonal elements of the  matrix \eqref{Gmat} are directly equal to the values of the QFI in given directions, which finishes the proof of Propositon \ref{propc}. 
\epf
The formula \eqref{qfifin} implies that given any finite dimensional system of a finite number of parties, and the collective Hamiltonian \eqref{genHcol}, the QFI averaged over any symmetric Hamiltonian distribution \eqref{HamSym} is equal to the QFI averaged over the basis elements of the space from which the Hamiltonian is sampled rescaled by a factor that depends solely on the way of sampling the Hamiltonian. The above was previously known only for the system of many qubits with collective Hamiltonian sampled from the unit sphere \cite{Hyllus12}.
\begin{remark}
The integral \eqref{qfifin} over the space of random Hamiltonians can be represented as an integral over their eigenvalues and eigenvectors. Due to the property \eqref{QFIpictures} it is equivalent to an integral over the orbit of isospectral states followed by integration over the spectrum $D$ of the Hamiltonian:
$$\int \QFI(\rho,\Ham_{\vec k})\diff \vec k =\int \QFI \left(U^{\otimes n}\rho (U^{\otimes n})^{\dagger},D\right)\diff U \diff D.$$
\end{remark}
Note that in our work we integrate over the set of random Hamiltonians $\Ham$, i.e. we average over the eigenvalues and eigenvectors of $\Ham$. In the approach of \cite{Oszmaniec2016} the average is performed over local orbits, which corresponds to the average over the eigenvectors only for a given fixed spectrum of the Hamiltonian.

\subsection{Mean Quantum Fisher Information for non-collective local random Hamiltonians of the form \eqref{genHncol}}
\label{sec:noncol}
\bpr
For an arbitrary non-collective Hamiltonian $\Ham_{\vec k_1,\ldots,\vec k_n}$ of the form  \eqref{genHncol}, in which local Hamiltonians $H_{\vec k_i}$, given by \eqref{localHam}, are sampled according to the measure $\diff\vec k_i$ fulfilling the conditions \eqref{HamSym}, the mean $\QFI$  reads:
\be
&&\int \QFI(\rho,\Ham_{\vec k_1,\ldots,\vec k_n})\diff \vec k_1\ldots\diff \vec k_n=\nonumber\\
&&\left(\frac{\int\Tr(H_{\vec k}^2)\diff\vec k}{\Tr(H_1^2)}\right)\left(\frac{\Tr(H_0)}{\Tr(H_1^2)}\right)^{2(n-1)}\frac{1}{r}\sum_{\pi}\sum_{i=1}^r\QFI(\rho,\Ham_{\pi(i)}),\nonumber\\
\label{thirdstepnc}
\ee
where:
\be
\Ham_{\pi(i)}=H_{i_1}\otimes H_{i_2}\otimes\ldots\otimes H_{i_n},
\ee
for $\{i_s\}_{s=1}^n=\pi(1,0,\ldots,0)$. 
\label{propnc}
\epr
In this case we proceed in full analogy to the previous one, however, due to a weaker symmetry of the global Hamiltonian \eqref{genHncol}  some steps  require more effort. The detailed technical proof is postponed to Appendix \ref{ProofProp2}.
The final result is similar to the previous one \eqref{qfifin}, but the average QFI is calculated in a different way. Instead of the average over collective Hamiltonians, in which all local Hamiltonians point in the same direction, we have the average over all possible single particle Hamiltonians extended trivially on the other parties.

\section{Characterization of states according to a sensitivity to local random unitary noise}

\subsection{Average QFI for pure states in terms of correlation tensors}
\label{pures}
In the case of pure states the analysis of a sensitivity of quantum states to  random noise can be simplified if we express  mean QFI from formulas \eqref{qfifin} and \eqref{thirdstepnc} in terms of correlation tensors of the states. Let us 
fix some local Hermitian basis consisting of traceless operators $\{H_i\}_{i=1}^d$, extended by the identity, denoted conveniently by $H_0$. Then any $n$-partite quantum state of $d$-level systems can be presented in a tensor form:
\be
\rho=\sum_{i_1,\ldots,i_n=0}^d T_{i_1i_2\ldots i_n}H_{i_1}\otimes H_{i_2}\otimes\ldots\otimes H_{i_n},\nonumber\\
\label{ctensor}
\ee
where the coefficients:
\be
T_{i_1i_2\ldots i_n}=\Tr(\rho\Ham_{i_1i_2\ldots i_n})
\ee
form the so called correlation tensor of a quantum state. Note that this definition of a correlation tensor differs slightly from the one conventionally used in the theory of entanglement \cite{Badziag08, Laskowski11, Laskowski13}, where the tensor basis in \eqref{ctensor} is normalized such that $T_{i_1i_2\ldots i_n}\in [-1,1]$ can be interpreted as a normalized correlation function. On the contrary, here the correlation tensor is always defined with respect to the same local bases, which are used to define the Hamiltonians \eqref{localHam} and \eqref{localHamnc}.

Since in the case of pure input states the QFI equals to four times the variance of the Hamiltonian \cite{Braunstein94}:
\be
\QFI(\rho_{\psi},H)=4 \left(\Tr(\rho H^2)-(\Tr(\rho H))^2\right), 
\ee
we can easily express both averages of the QFI \eqref{qfifin} and \eqref{thirdstepnc} in terms of the correlation tensor of an input state.
\bpr
The mean QFI for a pure input state with respect to a collective Hamiltonian \eqref{qfifin} is a function of local Bloch vectors and bipartite correlations of the state only:
\be
\sum_{i=1}^r \QFI(\rho_{\psi},\Ham_i)&=&\frac{4nr}{d}\Tr(H_1^2)+8\sum_{i=1}^r\sum_{\pi}T_{\pi(i,i,0,\ldots,0)}\nonumber\\
&-&4\sum_{i=1}^r\left(\sum_{\pi}T_{\pi(i,0,\ldots,0)}\right)^2.\nonumber\\
\ee
\label{propct}
\epr
\begin{proof}
See Appendix \ref{ProofProp3}.
\end{proof}
\begin{remark}
Note that the above mean QFI is constant for any states, which have vanishing all local Bloch vectors and bipartite correlations, which are known as \emph{2-uniform} states \cite{Scott04}.
\end{remark}
\bpr
The mean QFI for a pure input state with respect to a non-collective Hamiltonian \eqref{thirdstepnc} is a function of local Bloch vectors of the state only:
\be
&&\sum_{\pi}\sum_{i=1}^r\QFI(\rho,\Ham_{\pi(i)})=\frac{4nr}{d}\Tr(H_1^2)-4\sum_{\pi}\sum_{i=1}^r T_{\pi(i,0,\ldots,0)}^2.\nonumber\\
\ee
\label{propctn}
\epr
\begin{proof}
See Appendix \ref{ProofProp4}.
\end{proof}
\begin{remark}
Note that in this case the mean QFI is constant for all states with vanishing local Bloch vectors, known as \emph{1-uniform} states \cite{Scott04}.
\end{remark}

\subsection{The role of sampling random Hamiltonians}
\label{SecDist}

Both bounds on the fidelity of a state transfer \eqref{qfifin} and \eqref{thirdstepnc} depend on the method of sampling local random Hamiltonians via the mean value of a purity of a  Hamiltonian $\int \Tr(H_{\vec k}^2)\diff \vec k=E(\Tr(H^2))$.
Here we consider some special cases of random Hamiltonians of size $r$: sampled uniformly from different spheres and sampled entrywise from a normal distribution (Gaussian matrix ensambles). 

Firstly let us consider the Hamiltonian sampled uniformly from a 2-dimensional sphere $S^2$, which can be interpreted as the angular momentum operator pointing into a random direction:
\be
H_{\vec k}=\vec k\cdot\vec J=\sum_{i=1}^3 k_i J_i,
\label{HamJ}
\ee
in which $\vec k\in S^2$ is a normalized random vector, and the operators $\{J_x,J_y,J_z\}$ are the spin-$j$ angular momentum operators for $j=(d-1)/2$.

The above model can be generalized to the case $r=d^2-1$ by a uniform sampling from  $S^{d^2-2}$,:
\be
H_{\vec k}=\vec k\cdot\vec H=\sum_{i=1}^{d^2-1} k_i H_i,
\label{HamGL}
\ee
in which $\vec k\in S^{d^2-2}$ is a normalized vector sampled uniformly from $(d^2-2)$-dimensional sphere, and the operators $\{H_i\}$ are traceless Hermitian basis operators fulfilling $\Tr(H_i H_j)=2 \delta_{ij}$.  We may imagine that the operators $\{H_i\}$ act in an artificial $(d^2-1)-$dimensional Euclidean space, and that the Hamiltonian \eqref{HamGL} corresponds to the generator of the evolution in direction $\vec k$. 

The mean purity of Hamiltonians \eqref{HamJ} and \eqref{HamGL} can be easily found using the following Lemma:

\ble
Let $H_{\vec k}=\sum_{i=1}^r k_i H_i$ be a decomposition of an Hermitian matrix $H_{\vec k}$ in an orthogonal Hermitian set $\{H_i\}_{i=1}^r$, with coefficients $k_i$ taken from $(r-1)$-dimensional sphere ($\vec k\in S^{r-1}$). Assume also that $\Tr(H_i^2)=\Tr(H_1^2)=\textrm{const}$ for all $i,j$. Then the following relation holds:
\be
\int \Tr(H_{\vec k}^2)\diff \vec k=\Tr(H_1^2),
\ee
in which $\diff \vec k$ denotes uniform measure on the sphere $S^{r-1}$.
\label{lemmaHam}
\ele
\begin{proof}
\be
&&\int \Tr(H_{\vec k}^2)\diff \vec k=\sum_{i,j=1}^r \Tr(H_iH_j)\int k_ik_j\diff\vec k\nonumber\\
&&=\sum_{i=1}^r \Tr(H_1^2)\int k_i^2\diff\vec k=\Tr(H_1^2)\int \sum_{i=1}^r k_i^2\diff \vec k=\Tr(H_1^2)\nonumber\\
\ee
\end{proof}

Using the fact, that $\Tr(J_i^2)=\frac{2}{3}(j+1)(j+\frac{1}{2})j$ for $i=1,2,3$, where $j(j+1)$ is the eigenvalue of the operator $J^2=\sum_i J_i^2$, \cite{Devanathan02}, we obtain the following results \eqref{HamJ},
\be
&&\int_{\vec k\in S^2} \Tr(H_{\vec k}^2)\diff\vec k=\frac{2}{3}(j+1)(j+\frac{1}{2})j,\nonumber\\
&&\int_{\vec k\in S^{d^2-2}} \Tr(H_{\vec k}^2)\diff\vec k=2.
\ee

Further let us discuss the Gaussian ensambles. First 
we assume, that the Hamiltonian is given by a random Hermitian matrix $H$ pertaining Gaussian 
Unitary Ensemble,
\begin{equation}
H = \left(
\begin{smallmatrix}
\xi_{11} & \frac{\xi_{12}+ i \eta_{12}}{\sqrt{2}} & \dots \\
\frac{\xi_{12} - i \eta_{12}}{\sqrt{2}} &  \xi_{22} & \dots \\
\vdots & \vdots &\ddots
\end{smallmatrix}
\right),
\label{HamGUE}
\end{equation}
where $\xi_{ij}, \eta_{ij}$ are independent identically distributed (i.i.d.) random normal variables,
with mean equal to 0 and variance equal to 1.
From the above form we see  that the expected trace is given by
\begin{equation}
E\left( \Tr(H^2)\right) = E \left(
\sum_{i=1}^r \xi_{i,i}^2  + 
\sum_{i \neq j}^r \frac{\xi_{i,i}^2 + \eta_{ij}^2}{2}
\right).
\end{equation}
Since the random variables $\xi$ and $\eta$ have unit variance, we get
\begin{equation}
E \left(\Tr(H^2)\right)  = r+r(r-1) =r^2.
\end{equation}
Similarly we can consider a random Hamiltonian $H$ pertaining Gaussian Orthogonal 
Ensemble, 
\begin{equation}
H = \left(
\begin{smallmatrix}
\sqrt{2} \xi_{11} & \xi_{12} & \dots \\
\xi_{12} & \sqrt{2} \xi_{22} & \dots \\
\vdots & \vdots &\ddots
\end{smallmatrix}
\right),
\end{equation}
where $\xi_{ij}$ are i.i.d. normal random variables, with mean equal to 0 and 
variance equal to 1.
Now we calculate the expected trace and get 
\begin{equation}
E \left(\Tr(H^2)\right) = 2 r+r(r-1) =r(r+1).
\end{equation}
Although Hamiltonians \eqref{HamJ} and \eqref{HamGL} look rather differently than \eqref{HamGUE}, it turns out that they can be obtained as a normalized GUE distributions. Detailed discussion of this fact is postponed to Appendix \ref{sec:generalizedGUE}. 

\subsection{Sensitivity of several classes of states to the local unitary noise}
\label{examples}
The main bound \eqref{MainBoundv1} and both propositions \ref{propc} and \ref{propnc} show that the fidelity of the state affected by the channels \eqref{genUS2} and \eqref{genUSd} depends on the  Quantum Fisher Information averaged over the space of random Hamiltonians. Indeed, the lower the mean QFI, the more robust given state is against the unitary noise generated by a random local Hamiltonian. 

In this section we discuss our results for a several genuinely entangled $4$-qubit, $6$-qubit, and $4$-qutrit states, which frequently appear in quantum information. In particular we discuss Absolutely Maximally Entangled (AME) states of $2 n$ parties, which are $n$-uniform, so any its reduction to $n$-partite state is maximally mixed \cite{Scott04, Helwig13}. This property can be undestood in another way, namely AME states are genuinely entangled $2n$-partite states, which do not have any correlations between less than $n$ parties. 

In our analysis we focus on the following particular examples of four- and six-partite states:
\begin{itemize}
	\item $n$-qubit Greenberger-Horne-Zeilinger (GHZ) states \cite{GHZ} for $n=4,6$:
	\be
\ket{GHZ^{2}_n}=\frac{1}{\sqrt{2}}(\ket{0}^{\otimes n}+\ket{1}^{\otimes n}),
\ee
\item $n$-qubit Dicke states \cite{Dicke54} with $e$ excitations, for $n=6$, defined as: 
\be
\ket{D_n^e}=\frac{1}{\sqrt{\binom{n}{e}}}\sum_k \pi_k \left(\ket{1}^{\otimes e}\otimes\ket{0}^{\otimes(n-e)}\right), 
\ee
in which  $\sum_k \pi_k \left(\cdot\right)$ denotes the sum over all possible permutations of zeros and ones in the ket.
\item six-qubit Absolutely Maximally Entangled (AME) state \cite{Borras07}:
\be
\ket{AME^2_6} =\frac{c_1\ket{000000}+\ldots+c_{64}\ket{111111}}{4\sqrt{2}}
\ee
which is a superposition of all basis states with the following coefficients:  
\begin{footnotesize}
\begin{align*}
C_k=&\{1,0,0,-1,0,1,-1,0,0,1,1,0,-1,0,0,-1,0,-1,1,0,1,0,0,\\&-1,-1,0,0,-1,0,-1,-1,0,0,-1,-1,0,-1,0,0,-1,-1,0,\\&0,1,0, 1,-1,0,-1,0,0,-1,0,1,1,0,0,-1,1,0,-1,0,0,1\}.
\end{align*}
\end{footnotesize}
\item four-qutrit GHZ state:
\be
\ket{GHZ^3_4}=\frac{1}{\sqrt{3}}\left(\ket{0000}+\ket{1111}+\ket{2222}\right).
\ee
\item four-qutrit AME state \cite{Helwig13, AME}:
\be
&&\ket{AME^3_4}=\frac{1}{3}(\ket{0000}+\ket{0112}+\ket{0221}+\ket{1011}\nonumber\\
&&+\ket{1120}+\ket{1202}+\ket{2022}+\ket{2101}+\ket{2210})
\ee
\item the family of $4$-qutrit Dicke states \cite{Dicke54}: 
\begin{small}
\begin{align}
\ket{Q_4^1}&=\frac{1}{2}\sum_{k=1}^4\pi_k\left(\ket{0001}\right),
\\\ket{Q_4^2}&=\frac{1}{2\sqrt{7}}\left(2\sum_{k=1}^6\pi_k\left(\ket{0011}\right)+\sum_{k=1}^4\pi_k\left(\ket{0002}\right)\right),
\\\ket{Q_4^3}&=\frac{1}{2\sqrt{7}}\left(2\sum_{k=1}^4\pi_k\left(\ket{0111}\right)+\sum_{k=1}^{12}\pi_k\left(\ket{0012}\right)\right),
\\\ket{Q_4^4}&=\frac{1}{\sqrt{70}}\left(4\ket{1111}+2\sum_{k=1}^{12}\pi_k\left(\ket{0112}\right)+\sum_{k=1}^{6}\pi_k\left(\ket{0022}\right)\right),
\end{align}
\end{small}
\end{itemize}

We choose a random local Hamiltonian to be sampled uniformly from a sphere \eqref{HamJ}. In the case of qubit states the Hamiltonian reads:
\be
H_{\vec k}=\frac{1}{2}(\vec k\cdot\vec \sigma), \,\,\vec k \in S^2,
\label{HSigma}
\ee
in which  $\vec\sigma=\{\sigma_x, \sigma_y, \sigma_z\}$ is a vector of Pauli matrices. In the case of qutrit states we consider two cases: sampling from a sphere $S^2$:
\be
H_{\vec k}=(\vec k\cdot\vec J), \,\,\vec k\in S^2,
\label{HSpin}
\ee
where $\vec J$ is a vector of spin-1 matrices $\{J_x,J_y,J_z\}$, and sampling from a sphere $S^7$:
\be
H_{\vec k}=(\vec k\cdot\vec \lambda), \,\,\vec k\in S^7,
\label{HSpinS7}
\ee
where $\vec \lambda$ is a vector of eight Gell-Mann matrices.

\begin{table}
\begin{center}
\begin{tabular}{cccccc}
\hline\hline
d & n & $H_i$ & state  & $\langle\QFI^{\textrm{col}}\rangle$ & $\langle\QFI^{\textrm{non-col}}\rangle$ \\ 
\hline\hline
2 & 4 & $\sigma_i$ &$GHZ_4^2$       & 8 & 4 \\
2 & 4 & $\sigma_i$ &$D_4^1$       & $20\slash 3\approx 6.67$ & $11\slash 3\approx 3.67$ \\
2 & 4 & $\sigma_i$ &$D_4^2$      & 8 & 4 \\
\hline
2 & 6 & $\sigma_i$ &$GHZ_6^2$       & 16 & 6 \\
2 & 6 & $\sigma_i$ &$AME_6^2$   & 6  & 6 \\
2 & 6 & $\sigma_i$ &$D_6^3$      & 16 & 6 \\
\hline
3 & 4 & $J_i$ &$GHZ_4^3$  & $64\slash 3\approx 21.33$ & $32\slash 3\approx 10.67$ \\
3 & 4 & $J_i$&$AME_4^3$    & $32\slash 3\approx 10.67$ & $32\slash 3\approx 10.67$ \\
3 & 4 & $J_i$&$Q_4^1$  & $44\slash 3\approx 14.67$ & $23\slash 3\approx 7.67$ \\
3 & 4 & $J_i$&$Q_4^2$  & $64\slash 3\approx 21.33$ & $28\slash 3\approx 9.33$ \\
3 & 4 & $J_i$&$Q_4^3$  & $76\slash 3\approx 25.33$ & $31\slash 3\approx 10.33$ \\
3 & 4 & $J_i$&$Q_4^4$  & $80\slash 3\approx 26.67$ & $32\slash 3\approx 10.67$ \\
\hline
3 & 4 & $\lambda_i$ &$GHZ_4^3$  & $56\slash 3\approx18.67$   & $32\slash 3\approx10.67$ \\
3 & 4 & $\lambda_i$&$AME_4^3$    &$32\slash 3\approx10.67$   & $32\slash 3\approx10.67$ \\
3 & 4 & $\lambda_i$&$Q_4^1$  &  14                         & $19\slash 2=9.5$ \\
3 & 4 & $\lambda_i$&$Q_4^2$  & $806\slash 49\approx16.45$ & $991\slash 98\approx10.11$ \\
3 & 4 & $\lambda_i$&$Q_4^3$  &$842\slash 49\approx17.18$ & $1009\slash 98\approx10.30$ \\
3 & 4 & $\lambda_i$&$Q_4^4$  & $848\slash 49\approx17.31$ & $506\slash 49\approx10.33$ \\
\hline\hline
\end{tabular}
\end{center}
\caption{The table presents values of mean QFI for collective ($\langle\QFI^{\textrm{col}}\rangle$) and non-collective random Hamiltonians ($\langle\QFI^{\textrm{non-col}}\rangle$) for several four- and six-partite states. The columns $d$ and $n$ denote the number of levels and number of particles respectively. The column $H_i$ refers to a choice of a local basis for the random Hamiltonian ($\sigma_i$ -- Pauli matrices;  $J_i$ -- spin-j matrices with $j=(d-1)/2$; $\lambda_i$ -- Gell-Mann matrices).}
\label{TabMeanQFI}
\end{table}

In Table \ref{TabMeanQFI} we present values of the mean QFI for the above states for both ways of sampling local Hamiltonians \eqref{genUS2} and \eqref{genUSd}. It can be seen that for almost all the above defined classes of states the collective noise is more destructive than the non-collective one. In the case of AME states the influence of both types of noise is the same, which is a consequence of the fact, that they have vanishing local Bloch vectors and bipartite correlations, therefore the mean QFI of these states depends only on the dimension-dependent constant terms. 

\begin{figure}
\centering
\includegraphics[width=0.46\textwidth]{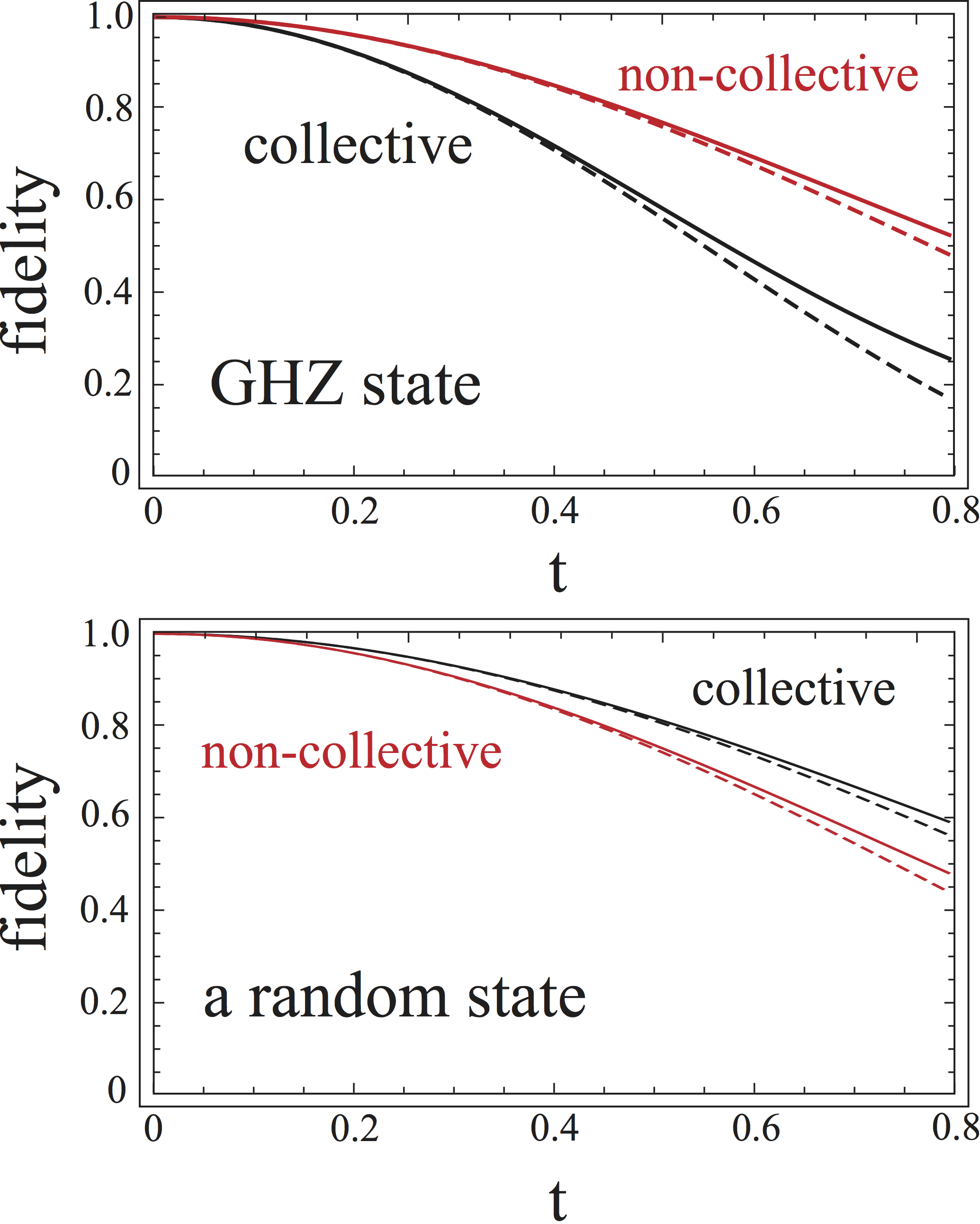}
\caption{Fidelity  between an initial and a final state, as a function of the interaction time $t$ \eqref{qfifidas}, for which approximation \eqref{MainBoundv1} holds. The time $t$ is defined in natural units. Figures represent comparison of fidelities for four-qubit GHZ state and a random pure state in the case of collective and non-collective local random noise. The solid lines represent the exact value of the fidelity, whereas the dashed lines represent the lower bound.}
\label{fig1}
\end{figure}

The natural conjecture may arise that the collective noise is always worse or has the same influence on the quantum states. However, we verified numerically that this is not true. We checked, that for a random pure state of four qubits, sampled according to a Haar measure the situation is inverse --- the non-collective noise is more destructive than the collective one (see Fig. \ref{fig1}).
The interesting fact about AME states is that in the case of a collective noise they are always the most robust, which suggests that from among the above classes of pure states they mostly resemble singlet states (which are completeley invariant with respect to any local unitary evolution). On the other hand in the case of a non-collective noise their robustness is as bad as in the case of a GHZ states.

\begin{figure}
\centering
\includegraphics[width=0.46\textwidth]{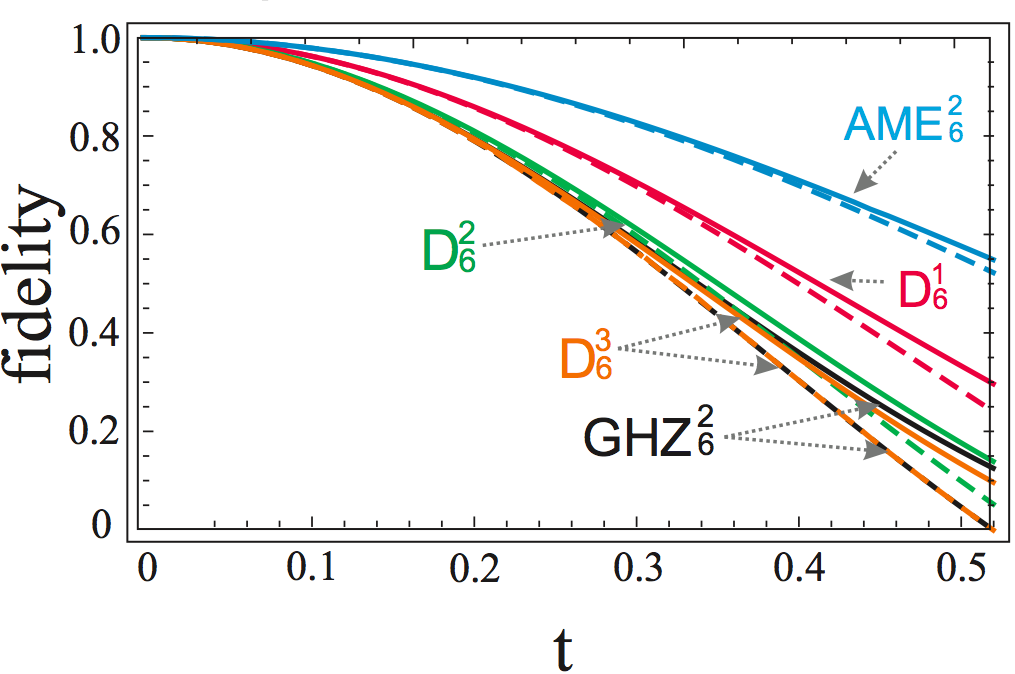}
\caption{Comparison of fidelities as functions of the interaction time $t$  \eqref{qfifidas} (in natural units) for several six-qubit  states in the case of collective local random noise. The solid lines represent the exact value of the fidelity, whereas the dashed lines represent the lower bound.}
\label{fig2}
\end{figure}

\begin{figure}
\centering
\includegraphics[width=0.46\textwidth]{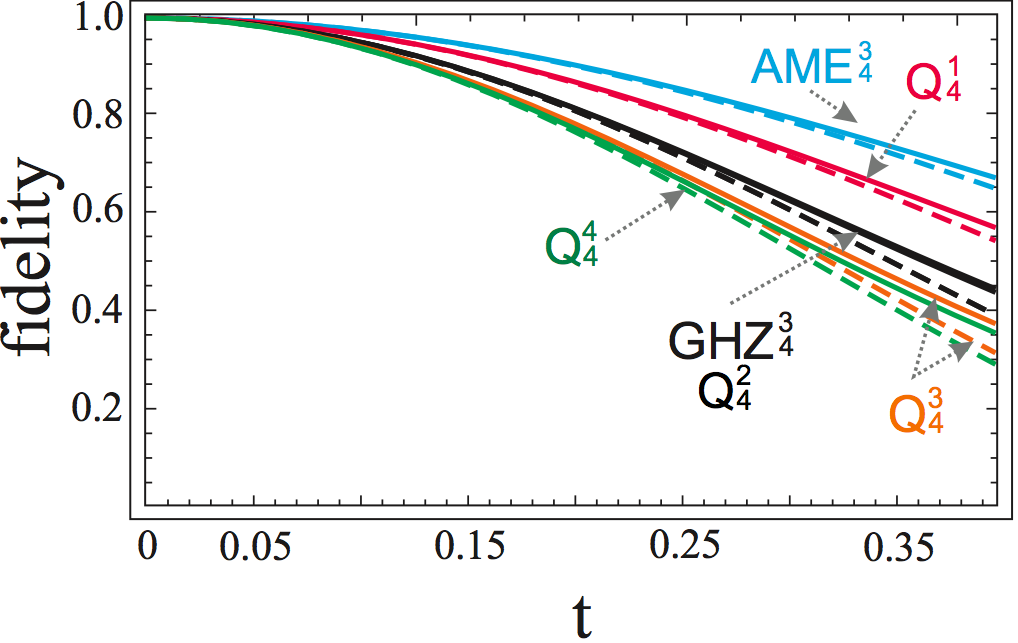}
\caption{As in Fig. \ref{fig2} for several four-qutrit  states.}
\label{fig3}
\end{figure}

The quality of  our approximation of the fidelity under local random noise for several classes of six-qubit and four-qutrit states is presented in Figures \ref{fig2} and \ref{fig3}.

More detailed numerical analysis in the case of collective evolution shows the following:
\begin{itemize}
	\item the relative error of the bound \eqref{MainBoundv1} is no greater than $1\%$ for at least $76.67\%$ of the allowed time range $t^*$ \eqref{qfifidas} of the approximation in the case of $\ket{AME^2_6}$ and $74.22\%$ of the time range in the case of $\ket{AME^3_4}$.
	\item taking into account the allowed time range $t^*$ \eqref{qfifidas}, the approximation is getting better with increasing number of particles, but while comparing accuracy in a fixed moment of time it is opposite. 
\end{itemize}

\section{Conclusions}
We showed that the influence of a local unitary noise generated by a random Hamiltonian can be characterized in terms of mean Quantum Fisher Information. Furthermore we showed that in the case of pure states the mean QFI is determined by lower order correlations (single particle and bipartite only). This fact leads us to the conclusion, that among states which are genuinely entangled, the more correlations are "stored" in the higher-order correlation sector, the more the state is insensitive to the local random noise.

Another interesting feature of the discussed class of states becomes clear if we compare collective and non-collective noise. States often used in the theory of quantum information can be considered as strongly non-typical. These states are more fragile with respect to the collective random noise, in contrast to generic random pure states, which suffer more under the action of a non-collective noise. 

From the point of view of the problems of random dynamics, our research demonstrates the difference between statical (uniform average over a symmetry group) and dynamical approach, in which we average over generators of evolution for a given evolution time treated as a parameter. Moreover we showed, that within the dynamical approach, the impact of the noisy random channel is actually identical for any distribution of the random Hamiltonians which fulfill very general symmetry conditions. 

Apart from the scenario of transmitting quantum states through noisy channels, our results on averaging the Quantum Fisher Information may have an impact in the fields of quantum metrology and entanglement detection. In the first case, the mean QFI determines the lower bound on the precision of estimation of an unknown parameter $t$ \cite{Helstrom67, FisherDef}: 
\be
\Delta t\geq \frac{1}{\sqrt{\langle\QFI(\rho,\Ham)\rangle}},
\ee
when a given quantum state $\rho$ evolves according to a random evolution determined by the distribution of the Hamiltonian $\Ham$ \cite{FisherDef}. In the second case, our result may stimulate the research on extending multiqubit entanglement criteria based on mean QFI \cite{Krischek11, Hyllus12, Toth12} to the case of entanglement between higher dimensional systems. 

\section{Acknowledgements}

We thank Rafa\l{} Demkowicz-Dobrza\'nski for helpful discussions and Micha\l{} Oszmaniec for useful correspondence. 

This research is supported by National Science Centre, Poland,
under grants number  2015/16/S/ST2/00447 within the project FUGA 4 for postdoctoral training (MM), 2014/14/M/ST2/00818 (AR, WL),   2015/17/B/ST6/01872 (ZP),
2011/02/A/ST1/00119 (KZ) and by Foundation for Polish Science
under program Start (MM). 
\appendix
\section{Some remarks on relation between Gaussian matrix ensambles and sampling from spheres} 
\label{sec:generalizedGUE}
\begin{remark}
Let $\{\Gamma_j\}_{i=1}^{d^2}$ be an orthonormal basis of the real space of 
Hermitian matrices, 
of size $d$, then for a vector $m=(m_i)_{i=1}^{d^2}$ of i.i.d standard, real 
normal variables, we have, that Hermitian operator $X$
\begin{equation}
X = \sum m_i \Gamma_i
\end{equation}
has the distribution of Gaussian Unitary Ensamble (GUE).
\end{remark}
The above remark follows from the definition of GUE and invariance of normal 
distribution with respect to orthogonal transformations. 
Note, that if one considers an orthogonal subset of Hermitian 
matrices $\Gamma_i$, which is not necessary a complete basis, then one gets a 
Gaussian distribution only on a subspace, spanned by $\Gamma_i$. This 
distribution is a projection of GUE distribution onto this subspace.

Consider the case of a random combination of $r$ Hermitian matrices $\Gamma_i$,
\begin{equation}
X = \sum_{i=1}^r s_i \Gamma_i,
\end{equation}
where the vector $s$ is a random  point on an appropriate real sphere sampled according to the uniform measure. The 
above vector can be obtained using normal variables, 
\begin{equation}
X \equiv \sum_{i=1}^r \frac{m_i}{\sqrt{\sum_{j=i}^r m_j^2}} \Gamma_i,
\end{equation}
where vector $m$ is a collection of standard i.i.d, real normal numbers.
From above consideration, we get that random combination of orthonormal 
Hermitian matrices, with uniform spherical weighs, can be obtained as a 
normalization of projected GUE random matrix. The projection is onto the 
$r$-dimensional subspace spanned by $\{\Gamma_i \}_{i=1}^r$.
\section{Random Hamiltonian evolution of a $5$ qubit GHZ}
\label{GHZ5App}
The coefficients \eqref{evolvedGHZ} of the five-qubit $\ket{GHZ^+_5}$ state \eqref{GHZdef} transformed by the evolution \eqref{genUS2} are given by:
\be
\zeta_1(t)&=&\frac{1}{21}\sin^2\left(\frac{t}{2}\right)(13+14\cos(t)+6\cos(2t)+2\cos(3t)),\nonumber\\
\zeta_2(t)&=&\frac{8}{63}\sin^4\left(\frac{t}{2}\right)(9+10\cos(t)+2\cos(2t)),\nonumber
\ee
\be
\zeta_3(t)&=&\frac{1}{1386}(382 + 302 \cos(t) + 302 \cos(2t) + 137 \cos(3t)\nonumber\\
 &+& 137 \cos(4t) + 126 \cos(5t)),\nonumber\\
\zeta_4(t)&=&\frac{1}{1386}(256 + 50 \cos(t) + 50 \cos(2t) - 115 \cos(3t)\nonumber\\
 &-& 115 \cos(4t) - 126 \cos(5t)),\nonumber
\ee
The above formulas imply, that all four coefficients $\zeta_i(t)$ cannot be simultaneously equal for any time instant $t$.

\section{Proof of the Lemma \ref{lem1}.}
\label{applemma}
First note that changing a sign in a decomposition of a random Hermitian matrix 
in some orthogonal basis is an orthogonal transformation on a real space of 
Hermitian matrices, hence for $i \neq j$ we have:
\be
&&\int \Tr(H_{\vec k}H_i)\Tr(H_{\vec k}H_j) \diff \vec k \nonumber\\
&&- \int \Tr(H_{\vec k}H_i)\Tr(H_{\vec k}H_j) \diff \vec k = 0,
\ee
which implies that:
\be
&&\int \Tr(H_{\vec k}H_i)\Tr(H_{\vec k}H_j) \diff \vec k= \delta_{ij}\int\left(Tr(H_{\vec k}H_i)\right)^2\diff\vec k.\nonumber\\
\label{lemmamain}
\ee
Now let us introduce an orthonormalized local Hermitian basis $\gamma_i=c_{\gamma}H_i$, where $c_{\gamma}=(\Tr(H_1^2))^{-\frac{1}{2}}$, such that $\Tr(\gamma_i\gamma_j)=\delta_{ij}$. Then we have:
\be
&&\int \left(\Tr(H_{\vec k}^2)\right)\diff\vec k=\nonumber\\
&&\int \Tr\left(\sum_{i=1}^r \Tr\left(H_{\vec k}\gamma_i \right) \gamma_i\sum_{j=1}^r\left(H_{\vec k}\gamma_j\right)\gamma_j\right)\diff \vec k=\nonumber\\
&&\int \sum_{i,j=1}^r \Tr\left(H_{\vec k}\gamma_i \right)\Tr\left(H_{\vec k}\gamma_j \right)\Tr(\gamma_i\gamma_j) \diff \vec k=\nonumber\\
&&\int \sum_{i=1}^r \Tr\left(H_{\vec k}\gamma_i \right)^2\diff\vec k=\sum_{i=1}^r\int\Tr\left(H_{\vec k}\gamma_i \right)^2\diff\vec k=\nonumber\\
&&r c_{\gamma}^2 \int\Tr\left(H_{\vec k}H_i \right)^2\diff\vec k,
\ee
where the last step follows from the invariance of the random Hamiltonian with respect to a permutation of coefficients. Therefore we have:
\be
&&\int\Tr\left(H_{\vec k}H_i \right)^2\diff\vec k=\frac{\Tr(H_1^2)}{r}\int\Tr(H_{\vec k}^2)\diff\vec k,
\label{TrHform}
\ee
which after putting into \eqref{lemmamain} proves the lemma.

\section{Proof of Proposition \ref{propnc}}
\label{ProofProp2}
We start with the direct definition of the QFI \eqref{qfidef}:
\be
&&\int \QFI(\rho,\Ham_{\vec k_1\ldots\vec k_n})\diff \vec k_1\ldots\diff \vec k_n\nonumber\\
&&=\int\sum_{m,l}f_{ml}\bra{m}\Ham_{\vec k_1\ldots\vec k_n}\ket{l}\bra{l}\Ham_{\vec k_1\ldots\vec k_n}\ket{m}\diff \vec k_1\ldots\diff \vec k_n,\nonumber\\
\label{qfiintnc}
\ee
in which $f_{ml}=\frac{2(\lambda_m-\lambda_l)^2}{\lambda_m+\lambda_l}$. 
We assume that each local Hamiltonian $H_{\vec k_l}$ can be expanded in some orthogonal basis  $\{H_i\}_{i=0}^r$:
\be
H_{\vec k_l}=\sum_{i=0}^r \alpha_i H_i,
\label{localHamnc}
\ee
for which we assume that $\Tr(H_i^2)=\Tr(H_1^2)=\textrm{const}$. Moreover in the current case we assume that the term $H_0$ is proportional to the identity, whereas all other basis elements $H_i$ are traceless. Now comes the main difference with respect to the previous case. Namely, we cannot use the collective direct-sum form basis \eqref{gengammac}, since the Hamiltonian is not collective. Instead we use the product tensor basis:
\be
\hat \gamma_{i_1,\ldots,i_n} &=& c_{\hat\gamma}\left(H_{i_1}\otimes H_{i_2}\otimes\ldots\otimes H_{i_n}\right)\nonumber\\
&=&c_{\hat\gamma}\Ham_{i_1,\ldots,i_n},
\label{gengammanc}
\ee
on which we impose the orthonormality condition:
\be
\Tr(\hat\gamma_{i_1,\ldots,i_n}\hat\gamma_{j_1,\ldots,j_n})=\delta_{i_1j_1}\ldots\delta_{i_nj_n}.
\ee
Thus the normalization coefficient is now given by:
\be
c_{\hat\gamma}^2=\frac{1}{(\Tr(H_1^2))^n}.
\label{gammanormnc}
\ee
Using the above basis we obtain the analogue of \eqref{secondstep}:
\be
&&\int \QFI(\rho,\Ham_{\vec k_1,\ldots,\vec k_n})\diff \vec k_1\ldots\diff \vec k_n=\nonumber\\
&&c_{\hat\gamma}^2\sum_{i,j,m,l}f_{ml}\bra{m}H_{i_1}\otimes\ldots\otimes H_{i_n}\ket{l}\bra{l}H_{j_1}\otimes\ldots\otimes H_{j_n}\ket{m}\nonumber\\
&&\times\int \Tr(\Ham_{\vec k_1,\ldots,\vec k_n}\hat \gamma_{i_1,\ldots,i_n})\Tr(\Ham_{\vec k_1,\ldots,\vec k_n}\hat \gamma_{j_1,\ldots,j_n})\diff \vec k_1\ldots\diff \vec k_n.\nonumber\\
\label{secondstepnc}
\ee
The remaining integral is now a bit more complicated:
\be
&&\int \Tr(\Ham_{\vec k_1,\ldots,\vec k_n}\hat \gamma_{i_1,\ldots,i_n})\Tr(\Ham_{\vec k_1,\ldots,\vec k_n}\hat \gamma_{j_1,\ldots,j_n})\diff \vec k_1\ldots\diff \vec k_n=\nonumber\\
&&c_{\hat\gamma}^2\int \Tr(H_{\vec k_1}H_{i_1}\otimes H_{i_2}\otimes\ldots+H_{i_1}\otimes H_{\vec k_2}H_{i_2}\otimes\ldots+\ldots)\nonumber\\
&&\times \Tr(H_{\vec k_1}H_{j_1}\otimes H_{j_2}\otimes\ldots+H_{j_1}\otimes H_{\vec k_2}H_{j_2}\otimes\ldots+\ldots)\nonumber\\
&&\diff \vec k_1\ldots\diff \vec k_n.
\label{singleintnc1}
\ee
Note that in the above integral, due to the assumption that $H_i$ for $i\neq 0$ are traceless, the only terms which survive have the sequences $\{i_s\}_{s=1}^n$ and $\{j_s\}_{s=1}^n$ in the form of $\pi(1,0,\ldots,0)$, where $\pi$ denotes a permutation. Moreover, both the sequences must be equal for a given permutation, since otherwise the integral takes the form:
\be
\int \Tr(H_{\vec k_s}H_{i_s})\diff\vec k_s\int\Tr(H_{\vec k_s'}H_{i_s'})\diff\vec k_s'=0,
\ee 
and vanishes due to the symmetry assumptions \eqref{HamSym}. Using these two facts we obtain:
\be
&&\int \Tr(\Ham_{\vec k_1,\ldots,\vec k_n}\hat \gamma_{i_1,\ldots,i_n})\Tr(\Ham_{\vec k_1,\ldots,\vec k_n}\hat \gamma_{j_1,\ldots,j_n})\diff \vec k_1\ldots\diff \vec k_n\nonumber\\
&&=c_{\hat\gamma}^2 (\Tr(H_0))^{2(n-1)}\delta_{\{i_s\},\{\pi(1,0,\ldots,0)\}}\int(\Tr(H_{\vec k_s}H_{i_s}))^2\diff\vec k_s,\nonumber\\
\label{singleintnc2}
\ee
where by the shorthand notation $\delta_{\{i_s\},\{\pi(1,0,\ldots,0)\}}$ we mean that $\{i_s\}=\{j_s\}=\pi(1,0,\ldots,0)$.
Finally, we put the above result to the formula \eqref{secondstepnc}, and using \eqref{TrHform} from the Lemma, we get:
\be
&&\int \QFI(\rho,\Ham_{\vec k_1,\ldots,\vec k_n})\diff \vec k_1\ldots\diff \vec k_n=\nonumber\\
&&c_{\hat\gamma}^4 (\Tr(H_0))^{2(n-1)}\delta_{\{i_s\},\{\pi(1,0,\ldots,0)\}}\frac{\Tr(H_1^2)}{r}\int\Tr(H_{\vec k}^2)\diff\vec k\nonumber\\
&&\times\sum_{i,j,m,l}f_{ml}\bra{m}H_{i_1}\otimes\ldots\otimes H_{i_n}\ket{l}\bra{l}H_{j_1}\otimes\ldots\otimes H_{j_n}\ket{m}\nonumber\\
&&=\left(\frac{\int\Tr(H_{\vec k}^2)\diff\vec k}{\Tr(H_1^2)}\right)\left(\frac{\Tr(H_0)}{\Tr(H_1^2)}\right)^{2(n-1)}\frac{\delta_{\{i_s\},\{\pi(1,0,\ldots,0)\}}}{r}\times\nonumber\\
&&\sum_{i,j,m,l}f_{ml}\bra{m}H_{i_1}\otimes\ldots\otimes H_{i_n}\ket{l}\bra{l}H_{j_1}\otimes\ldots\otimes H_{j_n}\ket{m}.\nonumber\\
\label{thirdstepncApp}
\ee
Introducing notation:
\be
\Ham_{\pi(i)}=H_{i_1}\otimes H_{i_2}\otimes\ldots\otimes H_{i_n},
\ee
for $\{i_s\}_{s=1}^n=\pi(1,0,\ldots,0)$, we arrive at the desired equation \eqref{thirdstepnc}:
\be
&&\int \QFI(\rho,\Ham_{\vec k_1,\ldots,\vec k_n})\diff \vec k_1\ldots\diff \vec k_n=\nonumber\\
&&\left(\frac{\int\Tr(H_{\vec k}^2)\diff\vec k}{\Tr(H_1^2)}\right)\left(\frac{\Tr(H_0)}{\Tr(H_1^2)}\right)^{2(n-1)}\frac{1}{r}\sum_{\pi}\sum_{i=1}^r\QFI(\rho,\Ham_{\pi(i)}),\nonumber\\
\label{thirdstepncApp1}
\ee

\section{Proof of Proposition \ref{propct}}
\label{ProofProp3}
\be
&&\sum_{i=1}^r \QFI(\rho_{\psi},\Ham_i)=4\sum_{i=1}^r\left(\Tr(\rho \Ham_i^2)-(\Tr(\rho \Ham_i))^2\right)=\nonumber\\
&&4\sum_{i=1}^r\left(2\Tr\left(\rho\sum_{\pi}\Ham_{\pi(i,i,0,\ldots,0)}\right)\right)+4\sum_{i=1}^r\frac{n}{d}\Tr(H_i^2)\nonumber\\
&&-4\sum_{i=1}^r\left(\Tr(\rho\Ham_i)\right)^2=8\sum_{i=1}^r\sum_{\pi}T_{\pi(i,i,0,\ldots,0)}\nonumber\\
&&+\frac{4nr}{d}\Tr(H_1^2)-4\sum_{i=1}^r\left(\sum_{\pi}T_{\pi(i,0,\ldots,0)}\right)^2.
\ee

\section{Proof of Proposition \ref{propctn}}
\label{ProofProp4}
\be
&&\sum_{\pi}\sum_{i=1}^r\QFI(\rho,\Ham_{\pi(i)})\nonumber\\
&&=4\sum_{\pi}\sum_{i=1}^r\Tr(\rho(H_{\pi(i)}\otimes H_{\pi(0)}\otimes\ldots\otimes H_{\pi(0)})^2)\nonumber\\
&&-4\sum_{\pi}\sum_{i=1}^r\Tr(\rho(H_{\pi(i)}\otimes H_{\pi(0)}\otimes\ldots\otimes H_{\pi(0)}))^2\nonumber\\
&&=4\sum_{\pi}\sum_{i=1}^r\Tr(\rho(H_{\pi(i)}^2\otimes H_{\pi(0)}^2\otimes\ldots\otimes H_{\pi(0)}^2))\nonumber\\
&&-4\sum_{\pi}\sum_{i=1}^r T_{\pi(i,0,\ldots,0)}^2,\nonumber\\
\ee

\bibliographystyle{apsrev4-1}

\begin{thebibliography}{49}%
\makeatletter
\providecommand \@ifxundefined [1]{%
 \@ifx{#1\undefined}
}%
\providecommand \@ifnum [1]{%
 \ifnum #1\expandafter \@firstoftwo
 \else \expandafter \@secondoftwo
 \fi
}%
\providecommand \@ifx [1]{%
 \ifx #1\expandafter \@firstoftwo
 \else \expandafter \@secondoftwo
 \fi
}%
\providecommand \natexlab [1]{#1}%
\providecommand \enquote  [1]{``#1''}%
\providecommand \bibnamefont  [1]{#1}%
\providecommand \bibfnamefont [1]{#1}%
\providecommand \citenamefont [1]{#1}%
\providecommand \href@noop [0]{\@secondoftwo}%
\providecommand \href [0]{\begingroup \@sanitize@url \@href}%
\providecommand \@href[1]{\@@startlink{#1}\@@href}%
\providecommand \@@href[1]{\endgroup#1\@@endlink}%
\providecommand \@sanitize@url [0]{\catcode `\\12\catcode `\$12\catcode
  `\&12\catcode `\#12\catcode `\^12\catcode `\_12\catcode `\%12\relax}%
\providecommand \@@startlink[1]{}%
\providecommand \@@endlink[0]{}%
\providecommand \url  [0]{\begingroup\@sanitize@url \@url }%
\providecommand \@url [1]{\endgroup\@href {#1}{\urlprefix }}%
\providecommand \urlprefix  [0]{URL }%
\providecommand \Eprint [0]{\href }%
\providecommand \doibase [0]{http://dx.doi.org/}%
\providecommand \selectlanguage [0]{\@gobble}%
\providecommand \bibinfo  [0]{\@secondoftwo}%
\providecommand \bibfield  [0]{\@secondoftwo}%
\providecommand \translation [1]{[#1]}%
\providecommand \BibitemOpen [0]{}%
\providecommand \bibitemStop [0]{}%
\providecommand \bibitemNoStop [0]{.\EOS\space}%
\providecommand \EOS [0]{\spacefactor3000\relax}%
\providecommand \BibitemShut  [1]{\csname bibitem#1\endcsname}%
\let\auto@bib@innerbib\@empty
\bibitem [{\citenamefont {Gisin}\ and\ \citenamefont {Thew}(2007)}]{Gisin07}%
  \BibitemOpen
  \bibfield  {author} {\bibinfo {author} {\bibfnamefont {N.}~\bibnamefont
  {Gisin}}\ and\ \bibinfo {author} {\bibfnamefont {R.}~\bibnamefont {Thew}},\
  }\href {\doibase doi:10.1038/nphoton.2007.22} {\bibfield  {journal} {\bibinfo
   {journal} {Nature Photonics}\ }\textbf {\bibinfo {volume} {1}},\ \bibinfo
  {pages} {165} (\bibinfo {year} {2007})}\BibitemShut {NoStop}%
\bibitem [{\citenamefont {Wang}(2005)}]{Wang05}%
  \BibitemOpen
  \bibfield  {author} {\bibinfo {author} {\bibfnamefont {X.-B.}\ \bibnamefont
  {Wang}},\ }\href {\doibase 10.1103/PhysRevA.72.050304} {\bibfield  {journal}
  {\bibinfo  {journal} {Phys. Rev. A}\ }\textbf {\bibinfo {volume} {72}},\
  \bibinfo {pages} {050304} (\bibinfo {year} {2005})}\BibitemShut {NoStop}%
\bibitem [{\citenamefont {Emerson}\ \emph {et~al.}(2005)\citenamefont
  {Emerson}, \citenamefont {Alicki},\ and\ \citenamefont
  {\.Zyczkowski}}]{Emerson05}%
  \BibitemOpen
  \bibfield  {author} {\bibinfo {author} {\bibfnamefont {J.}~\bibnamefont
  {Emerson}}, \bibinfo {author} {\bibfnamefont {R.}~\bibnamefont {Alicki}}, \
  and\ \bibinfo {author} {\bibfnamefont {K.}~\bibnamefont {\.Zyczkowski}},\
  }\href@noop {} {\bibfield  {journal} {\bibinfo  {journal} {J. Opt. B: Quantum
  Semiclass. Opt.}\ }\textbf {\bibinfo {volume} {7}},\ \bibinfo {pages} {S347}
  (\bibinfo {year} {2005})}\BibitemShut {NoStop}%
\bibitem [{\citenamefont {Bartlett}\ \emph {et~al.}(2007)\citenamefont
  {Bartlett}, \citenamefont {Rudolph},\ and\ \citenamefont
  {Spekkens}}]{Bartlett07}%
  \BibitemOpen
  \bibfield  {author} {\bibinfo {author} {\bibfnamefont {S.~D.}\ \bibnamefont
  {Bartlett}}, \bibinfo {author} {\bibfnamefont {T.}~\bibnamefont {Rudolph}}, \
  and\ \bibinfo {author} {\bibfnamefont {R.~W.}\ \bibnamefont {Spekkens}},\
  }\href {\doibase 10.1103/RevModPhys.79.555} {\bibfield  {journal} {\bibinfo
  {journal} {Rev. Mod. Phys.}\ }\textbf {\bibinfo {volume} {79}},\ \bibinfo
  {pages} {555} (\bibinfo {year} {2007})}\BibitemShut {NoStop}%
\bibitem [{\citenamefont {Vinayak}\ and\ \citenamefont
  {\v{Z}nidari\v{c}}(2012)}]{RandHam12}%
  \BibitemOpen
  \bibfield  {author} {\bibinfo {author} {\bibnamefont {Vinayak}}\ and\
  \bibinfo {author} {\bibfnamefont {M.}~\bibnamefont {\v{Z}nidari\v{c}}},\
  }\href {http://stacks.iop.org/1751-8121/45/i=12/a=125204} {\bibfield
  {journal} {\bibinfo  {journal} {J. Phys. A: Math. Theor.}\ }\textbf {\bibinfo
  {volume} {45}},\ \bibinfo {pages} {125204} (\bibinfo {year}
  {2012})}\BibitemShut {NoStop}%
\bibitem [{\citenamefont {\ifmmode \check{Z}\else
  \v{Z}\fi{}nidari\ifmmode~\check{c}\else \v{c}\fi{}}\ \emph
  {et~al.}(2011)\citenamefont {\ifmmode \check{Z}\else
  \v{Z}\fi{}nidari\ifmmode~\check{c}\else \v{c}\fi{}}, \citenamefont {Pineda},\
  and\ \citenamefont {Garc\'{\i}a-Mata}}]{Znidaric2011}%
  \BibitemOpen
  \bibfield  {author} {\bibinfo {author} {\bibfnamefont {M.}~\bibnamefont
  {\ifmmode \check{Z}\else \v{Z}\fi{}nidari\ifmmode~\check{c}\else
  \v{c}\fi{}}}, \bibinfo {author} {\bibfnamefont {C.}~\bibnamefont {Pineda}}, \
  and\ \bibinfo {author} {\bibfnamefont {I.}~\bibnamefont {Garc\'{\i}a-Mata}},\
  }\href {\doibase 10.1103/PhysRevLett.107.080404} {\bibfield  {journal}
  {\bibinfo  {journal} {Phys. Rev. Lett.}\ }\textbf {\bibinfo {volume} {107}},\
  \bibinfo {pages} {080404} (\bibinfo {year} {2011})}\BibitemShut {NoStop}%
\bibitem [{\citenamefont {Brand\~ao}\ \emph {et~al.}(2012)\citenamefont
  {Brand\~ao}, \citenamefont {\ifmmode \acute{C}\else
  \'{C}\fi{}wikli\ifmmode~\acute{n}\else \'{n}\fi{}ski}, \citenamefont
  {Horodecki}, \citenamefont {Horodecki}, \citenamefont {Korbicz},\ and\
  \citenamefont {Mozrzymas}}]{Brandao2012}%
  \BibitemOpen
  \bibfield  {author} {\bibinfo {author} {\bibfnamefont {F.~G. S.~L.}\
  \bibnamefont {Brand\~ao}}, \bibinfo {author} {\bibfnamefont {P.}~\bibnamefont
  {\ifmmode \acute{C}\else \'{C}\fi{}wikli\ifmmode~\acute{n}\else
  \'{n}\fi{}ski}}, \bibinfo {author} {\bibfnamefont {M.}~\bibnamefont
  {Horodecki}}, \bibinfo {author} {\bibfnamefont {P.}~\bibnamefont
  {Horodecki}}, \bibinfo {author} {\bibfnamefont {J.~K.}\ \bibnamefont
  {Korbicz}}, \ and\ \bibinfo {author} {\bibfnamefont {M.}~\bibnamefont
  {Mozrzymas}},\ }\href {\doibase 10.1103/PhysRevE.86.031101} {\bibfield
  {journal} {\bibinfo  {journal} {Phys. Rev. E}\ }\textbf {\bibinfo {volume}
  {86}},\ \bibinfo {pages} {031101} (\bibinfo {year} {2012})}\BibitemShut
  {NoStop}%
\bibitem [{\citenamefont {Gessner}\ and\ \citenamefont
  {Breuer}(2013)}]{Gessner2013}%
  \BibitemOpen
  \bibfield  {author} {\bibinfo {author} {\bibfnamefont {M.}~\bibnamefont
  {Gessner}}\ and\ \bibinfo {author} {\bibfnamefont {H.-P.}\ \bibnamefont
  {Breuer}},\ }\href {\doibase 10.1103/PhysRevE.87.042128} {\bibfield
  {journal} {\bibinfo  {journal} {Phys. Rev. E}\ }\textbf {\bibinfo {volume}
  {87}},\ \bibinfo {pages} {042128} (\bibinfo {year} {2013})}\BibitemShut
  {NoStop}%
\bibitem [{\citenamefont {Bartlett}\ \emph {et~al.}(2003)\citenamefont
  {Bartlett}, \citenamefont {Rudolph},\ and\ \citenamefont
  {Spekkens}}]{Bartlett03}%
  \BibitemOpen
  \bibfield  {author} {\bibinfo {author} {\bibfnamefont {S.~D.}\ \bibnamefont
  {Bartlett}}, \bibinfo {author} {\bibfnamefont {T.}~\bibnamefont {Rudolph}}, \
  and\ \bibinfo {author} {\bibfnamefont {R.~W.}\ \bibnamefont {Spekkens}},\
  }\href {\doibase 10.1103/PhysRevLett.91.027901} {\bibfield  {journal}
  {\bibinfo  {journal} {Phys. Rev. Lett.}\ }\textbf {\bibinfo {volume} {91}},\
  \bibinfo {pages} {027901} (\bibinfo {year} {2003})}\BibitemShut {NoStop}%
\bibitem [{\citenamefont {Gour}\ \emph {et~al.}(2009)\citenamefont {Gour},
  \citenamefont {Marvian},\ and\ \citenamefont {Spekkens}}]{Gour09}%
  \BibitemOpen
  \bibfield  {author} {\bibinfo {author} {\bibfnamefont {G.}~\bibnamefont
  {Gour}}, \bibinfo {author} {\bibfnamefont {I.}~\bibnamefont {Marvian}}, \
  and\ \bibinfo {author} {\bibfnamefont {R.~W.}\ \bibnamefont {Spekkens}},\
  }\href {\doibase 10.1103/PhysRevA.80.012307} {\bibfield  {journal} {\bibinfo
  {journal} {Phys. Rev. A}\ }\textbf {\bibinfo {volume} {80}},\ \bibinfo
  {pages} {012307} (\bibinfo {year} {2009})}\BibitemShut {NoStop}%
\bibitem [{\citenamefont {Palmer}\ \emph {et~al.}(2014)\citenamefont {Palmer},
  \citenamefont {Girelli},\ and\ \citenamefont {Bartlett}}]{Bartlett14}%
  \BibitemOpen
  \bibfield  {author} {\bibinfo {author} {\bibfnamefont {M.~C.}\ \bibnamefont
  {Palmer}}, \bibinfo {author} {\bibfnamefont {F.}~\bibnamefont {Girelli}}, \
  and\ \bibinfo {author} {\bibfnamefont {S.~D.}\ \bibnamefont {Bartlett}},\
  }\href {\doibase 10.1103/PhysRevA.89.052121} {\bibfield  {journal} {\bibinfo
  {journal} {Phys. Rev. A}\ }\textbf {\bibinfo {volume} {89}},\ \bibinfo
  {pages} {052121} (\bibinfo {year} {2014})}\BibitemShut {NoStop}%
\bibitem [{\citenamefont {Werner}(1989)}]{Werner89}%
  \BibitemOpen
  \bibfield  {author} {\bibinfo {author} {\bibfnamefont {R.~F.}\ \bibnamefont
  {Werner}},\ }\href {\doibase 10.1103/PhysRevA.40.4277} {\bibfield  {journal}
  {\bibinfo  {journal} {Phys. Rev. A}\ }\textbf {\bibinfo {volume} {40}},\
  \bibinfo {pages} {4277} (\bibinfo {year} {1989})}\BibitemShut {NoStop}%
\bibitem [{\citenamefont {Horodecki}\ and\ \citenamefont
  {Horodecki}(1999)}]{HH99}%
  \BibitemOpen
  \bibfield  {author} {\bibinfo {author} {\bibfnamefont {M.}~\bibnamefont
  {Horodecki}}\ and\ \bibinfo {author} {\bibfnamefont {P.}~\bibnamefont
  {Horodecki}},\ }\href {\doibase 10.1103/PhysRevA.59.4206} {\bibfield
  {journal} {\bibinfo  {journal} {Phys. Rev. A}\ }\textbf {\bibinfo {volume}
  {59}},\ \bibinfo {pages} {4206} (\bibinfo {year} {1999})}\BibitemShut
  {NoStop}%
\bibitem [{\citenamefont {Johnson}\ and\ \citenamefont
  {Viola}(2013)}]{Johnson13}%
  \BibitemOpen
  \bibfield  {author} {\bibinfo {author} {\bibfnamefont {P.~D.}\ \bibnamefont
  {Johnson}}\ and\ \bibinfo {author} {\bibfnamefont {L.}~\bibnamefont
  {Viola}},\ }\href {\doibase 10.1103/PhysRevA.88.032323} {\bibfield  {journal}
  {\bibinfo  {journal} {Phys. Rev. A}\ }\textbf {\bibinfo {volume} {88}},\
  \bibinfo {pages} {032323} (\bibinfo {year} {2013})}\BibitemShut {NoStop}%
\bibitem [{\citenamefont {Helstrom}(1967)}]{Helstrom67}%
  \BibitemOpen
  \bibfield  {author} {\bibinfo {author} {\bibfnamefont {C.~W.}\ \bibnamefont
  {Helstrom}},\ }\href@noop {} {\bibfield  {journal} {\bibinfo  {journal}
  {Phys. Lett. A}\ }\textbf {\bibinfo {volume} {25}},\ \bibinfo {pages} {101}
  (\bibinfo {year} {1967})}\BibitemShut {NoStop}%
\bibitem [{\citenamefont {Braunstein}\ and\ \citenamefont
  {Caves}(1994)}]{Braunstein94}%
  \BibitemOpen
  \bibfield  {author} {\bibinfo {author} {\bibfnamefont {S.~L.}\ \bibnamefont
  {Braunstein}}\ and\ \bibinfo {author} {\bibfnamefont {C.~M.}\ \bibnamefont
  {Caves}},\ }\href@noop {} {\bibfield  {journal} {\bibinfo  {journal} {Phys.
  Rev. Lett.}\ }\textbf {\bibinfo {volume} {72}},\ \bibinfo {pages} {3439}
  (\bibinfo {year} {1994})}\BibitemShut {NoStop}%
\bibitem [{\citenamefont {T\'oth}\ and\ \citenamefont
  {Apellaniz}(2014)}]{FisherDef}%
  \BibitemOpen
  \bibfield  {author} {\bibinfo {author} {\bibfnamefont {G.}~\bibnamefont
  {T\'oth}}\ and\ \bibinfo {author} {\bibfnamefont {I.}~\bibnamefont
  {Apellaniz}},\ }\href@noop {} {\bibfield  {journal} {\bibinfo  {journal} {J.
  Phys. A: Math. Theor.}\ }\textbf {\bibinfo {volume} {47}},\ \bibinfo {pages}
  {424006} (\bibinfo {year} {2014})}\BibitemShut {NoStop}%
\bibitem [{\citenamefont {Song}\ \emph {et~al.}(2013)\citenamefont {Song},
  \citenamefont {Luo}, \citenamefont {Li},\ and\ \citenamefont
  {Chang}}]{clone1}%
  \BibitemOpen
  \bibfield  {author} {\bibinfo {author} {\bibfnamefont {H.}~\bibnamefont
  {Song}}, \bibinfo {author} {\bibfnamefont {S.}~\bibnamefont {Luo}}, \bibinfo
  {author} {\bibfnamefont {N.}~\bibnamefont {Li}}, \ and\ \bibinfo {author}
  {\bibfnamefont {L.}~\bibnamefont {Chang}},\ }\href {\doibase
  10.1103/PhysRevA.88.042121} {\bibfield  {journal} {\bibinfo  {journal} {Phys.
  Rev. A}\ }\textbf {\bibinfo {volume} {88}},\ \bibinfo {pages} {042121}
  (\bibinfo {year} {2013})}\BibitemShut {NoStop}%
\bibitem [{\citenamefont {Xiao}\ \emph {et~al.}(2014)\citenamefont {Xiao},
  \citenamefont {Yao}, \citenamefont {Zhou},\ and\ \citenamefont
  {Wang}}]{clone2}%
  \BibitemOpen
  \bibfield  {author} {\bibinfo {author} {\bibfnamefont {X.}~\bibnamefont
  {Xiao}}, \bibinfo {author} {\bibfnamefont {Y.}~\bibnamefont {Yao}}, \bibinfo
  {author} {\bibfnamefont {L.-M.}\ \bibnamefont {Zhou}}, \ and\ \bibinfo
  {author} {\bibfnamefont {X.}~\bibnamefont {Wang}},\ }\href@noop {} {\bibfield
   {journal} {\bibinfo  {journal} {Sci. Rep.}\ }\textbf {\bibinfo {volume}
  {4}},\ \bibinfo {pages} {7361} (\bibinfo {year} {2014})}\BibitemShut
  {NoStop}%
\bibitem [{\citenamefont {\v{S}afranek}\ \emph {et~al.}(2015)\citenamefont
  {\v{S}afranek}, \citenamefont {Ahmadi},\ and\ \citenamefont
  {Fuentes}}]{Ahmadi15}%
  \BibitemOpen
  \bibfield  {author} {\bibinfo {author} {\bibfnamefont {D.}~\bibnamefont
  {\v{S}afranek}}, \bibinfo {author} {\bibfnamefont {M.}~\bibnamefont
  {Ahmadi}}, \ and\ \bibinfo {author} {\bibfnamefont {I.}~\bibnamefont
  {Fuentes}},\ }\href@noop {} {\bibfield  {journal} {\bibinfo  {journal} {New
  J. Phys.}\ }\textbf {\bibinfo {volume} {17}},\ \bibinfo {pages} {033012}
  (\bibinfo {year} {2015})}\BibitemShut {NoStop}%
\bibitem [{\citenamefont {Demkowicz-Dobrza\ifmmode~\acute{n}\else
  \'{n}\fi{}ski}\ and\ \citenamefont {Markiewicz}(2015)}]{ourGrover}%
  \BibitemOpen
  \bibfield  {author} {\bibinfo {author} {\bibfnamefont {R.}~\bibnamefont
  {Demkowicz-Dobrza\ifmmode~\acute{n}\else \'{n}\fi{}ski}}\ and\ \bibinfo
  {author} {\bibfnamefont {M.}~\bibnamefont {Markiewicz}},\ }\href {\doibase
  10.1103/PhysRevA.91.062322} {\bibfield  {journal} {\bibinfo  {journal} {Phys.
  Rev. A}\ }\textbf {\bibinfo {volume} {91}},\ \bibinfo {pages} {062322}
  (\bibinfo {year} {2015})}\BibitemShut {NoStop}%
\bibitem [{\citenamefont {Oszmaniec}\ \emph {et~al.}(2016)\citenamefont
  {Oszmaniec}, \citenamefont {Augusiak}, \citenamefont {Gogolin}, \citenamefont
  {Ko\l{}ody\ifmmode~\acute{n}\else \'{n}\fi{}ski}, \citenamefont {Ac\'{\i}n},\
  and\ \citenamefont {Lewenstein}}]{Oszmaniec2016}%
  \BibitemOpen
  \bibfield  {author} {\bibinfo {author} {\bibfnamefont {M.}~\bibnamefont
  {Oszmaniec}}, \bibinfo {author} {\bibfnamefont {R.}~\bibnamefont {Augusiak}},
  \bibinfo {author} {\bibfnamefont {C.}~\bibnamefont {Gogolin}}, \bibinfo
  {author} {\bibfnamefont {J.}~\bibnamefont {Ko\l{}ody\ifmmode~\acute{n}\else
  \'{n}\fi{}ski}}, \bibinfo {author} {\bibfnamefont {A.}~\bibnamefont
  {Ac\'{\i}n}}, \ and\ \bibinfo {author} {\bibfnamefont {M.}~\bibnamefont
  {Lewenstein}},\ }\href {\doibase 10.1103/PhysRevX.6.041044} {\bibfield
  {journal} {\bibinfo  {journal} {Phys. Rev. X}\ }\textbf {\bibinfo {volume}
  {6}},\ \bibinfo {pages} {041044} (\bibinfo {year} {2016})}\BibitemShut
  {NoStop}%
\bibitem [{\citenamefont {Bures}(1969)}]{Bures69}%
  \BibitemOpen
  \bibfield  {author} {\bibinfo {author} {\bibfnamefont {D.}~\bibnamefont
  {Bures}},\ }\href@noop {} {\bibfield  {journal} {\bibinfo  {journal} {Trans.
  Am. Math. Soc.}\ }\textbf {\bibinfo {volume} {135}},\ \bibinfo {pages} {199}
  (\bibinfo {year} {1969})}\BibitemShut {NoStop}%
\bibitem [{\citenamefont {Jozsa}(1994)}]{Jozsa94}%
  \BibitemOpen
  \bibfield  {author} {\bibinfo {author} {\bibfnamefont {R.}~\bibnamefont
  {Jozsa}},\ }\href@noop {} {\bibfield  {journal} {\bibinfo  {journal} {J. Mod.
  Opt.}\ }\textbf {\bibinfo {volume} {41}},\ \bibinfo {pages} {2315} (\bibinfo
  {year} {1994})}\BibitemShut {NoStop}%
\bibitem [{\citenamefont {Bengtsson}\ and\ \citenamefont
  {\.Zyczkowski}(2006)}]{Bengtsson2006}%
  \BibitemOpen
  \bibfield  {author} {\bibinfo {author} {\bibfnamefont {I.}~\bibnamefont
  {Bengtsson}}\ and\ \bibinfo {author} {\bibfnamefont {K.}~\bibnamefont
  {\.Zyczkowski}},\ }\href@noop {} {\emph {\bibinfo {title} {Geometry of
  quantum states: an introduction to quantum entanglement}}}\ (\bibinfo
  {publisher} {Cambridge Univeristy Press},\ \bibinfo {year}
  {2006})\BibitemShut {NoStop}%
\bibitem [{\citenamefont {Taddei}\ \emph {et~al.}(2013)\citenamefont {Taddei},
  \citenamefont {Escher}, \citenamefont {Davidovich},\ and\ \citenamefont
  {de~Matos~Filho}}]{Taddei13}%
  \BibitemOpen
  \bibfield  {author} {\bibinfo {author} {\bibfnamefont {M.~M.}\ \bibnamefont
  {Taddei}}, \bibinfo {author} {\bibfnamefont {B.~M.}\ \bibnamefont {Escher}},
  \bibinfo {author} {\bibfnamefont {L.}~\bibnamefont {Davidovich}}, \ and\
  \bibinfo {author} {\bibfnamefont {R.~L.}\ \bibnamefont {de~Matos~Filho}},\
  }\href@noop {} {\bibfield  {journal} {\bibinfo  {journal} {Phys Rev. Lett.}\
  }\textbf {\bibinfo {volume} {110}},\ \bibinfo {pages} {050402} (\bibinfo
  {year} {2013})}\BibitemShut {NoStop}%
\bibitem [{\citenamefont {Mandelstam}\ and\ \citenamefont
  {Tamm}(1945)}]{Mandelstam45}%
  \BibitemOpen
  \bibfield  {author} {\bibinfo {author} {\bibfnamefont {L.}~\bibnamefont
  {Mandelstam}}\ and\ \bibinfo {author} {\bibfnamefont {I.}~\bibnamefont
  {Tamm}},\ }\href@noop {} {\bibfield  {journal} {\bibinfo  {journal} {J. Phys.
  (USSR)}\ }\textbf {\bibinfo {volume} {9}},\ \bibinfo {pages} {249} (\bibinfo
  {year} {1945})}\BibitemShut {NoStop}%
\bibitem [{\citenamefont {Fr\"owis}(2012)}]{Frowis12}%
  \BibitemOpen
  \bibfield  {author} {\bibinfo {author} {\bibfnamefont {F.}~\bibnamefont
  {Fr\"owis}},\ }\href {\doibase 10.1103/PhysRevA.85.052127} {\bibfield
  {journal} {\bibinfo  {journal} {Phys. Rev. A}\ }\textbf {\bibinfo {volume}
  {85}},\ \bibinfo {pages} {052127} (\bibinfo {year} {2012})}\BibitemShut
  {NoStop}%
\bibitem [{\citenamefont {Giovannetti}\ \emph {et~al.}(2003)\citenamefont
  {Giovannetti}, \citenamefont {Lloyd},\ and\ \citenamefont
  {Maccone}}]{Giovannetti03}%
  \BibitemOpen
  \bibfield  {author} {\bibinfo {author} {\bibfnamefont {V.}~\bibnamefont
  {Giovannetti}}, \bibinfo {author} {\bibfnamefont {S.}~\bibnamefont {Lloyd}},
  \ and\ \bibinfo {author} {\bibfnamefont {L.}~\bibnamefont {Maccone}},\
  }\href@noop {} {\bibfield  {journal} {\bibinfo  {journal} {Phys. Rev. A}\
  }\textbf {\bibinfo {volume} {67}},\ \bibinfo {pages} {052109} (\bibinfo
  {year} {2003})}\BibitemShut {NoStop}%
\bibitem [{\citenamefont {Pires}\ \emph {et~al.}(2016)\citenamefont {Pires},
  \citenamefont {Cianciaruso}, \citenamefont {C\'eleri}, \citenamefont
  {Adesso},\ and\ \citenamefont {Soares-Pinto}}]{Pires16}%
  \BibitemOpen
  \bibfield  {author} {\bibinfo {author} {\bibfnamefont {D.~P.}\ \bibnamefont
  {Pires}}, \bibinfo {author} {\bibfnamefont {M.}~\bibnamefont {Cianciaruso}},
  \bibinfo {author} {\bibfnamefont {L.~C.}\ \bibnamefont {C\'eleri}}, \bibinfo
  {author} {\bibfnamefont {G.}~\bibnamefont {Adesso}}, \ and\ \bibinfo {author}
  {\bibfnamefont {D.~O.}\ \bibnamefont {Soares-Pinto}},\ }\href {\doibase
  10.1103/PhysRevX.6.021031} {\bibfield  {journal} {\bibinfo  {journal} {Phys.
  Rev. X}\ }\textbf {\bibinfo {volume} {6}},\ \bibinfo {pages} {021031}
  (\bibinfo {year} {2016})}\BibitemShut {NoStop}%
\bibitem [{\citenamefont {Gibilisco}\ and\ \citenamefont
  {Isola}(2003)}]{WYdist}%
  \BibitemOpen
  \bibfield  {author} {\bibinfo {author} {\bibfnamefont {P.}~\bibnamefont
  {Gibilisco}}\ and\ \bibinfo {author} {\bibfnamefont {T.}~\bibnamefont
  {Isola}},\ }\href@noop {} {\bibfield  {journal} {\bibinfo  {journal} {J.
  Math. Phys.}\ }\textbf {\bibinfo {volume} {44}},\ \bibinfo {pages} {3752}
  (\bibinfo {year} {2003})}\BibitemShut {NoStop}%
\bibitem [{\citenamefont {Ma}\ \emph {et~al.}(2008)\citenamefont {Ma},
  \citenamefont {Zhang},\ and\ \citenamefont {Chen}}]{Afidelity}%
  \BibitemOpen
  \bibfield  {author} {\bibinfo {author} {\bibfnamefont {Z.}~\bibnamefont
  {Ma}}, \bibinfo {author} {\bibfnamefont {F.-L.}\ \bibnamefont {Zhang}}, \
  and\ \bibinfo {author} {\bibfnamefont {J.-L.}\ \bibnamefont {Chen}},\ }\href
  {\doibase 10.1103/PhysRevA.78.064305} {\bibfield  {journal} {\bibinfo
  {journal} {Phys. Rev. A}\ }\textbf {\bibinfo {volume} {78}},\ \bibinfo
  {pages} {064305} (\bibinfo {year} {2008})}\BibitemShut {NoStop}%
\bibitem [{\citenamefont {Hyllus}\ \emph {et~al.}(2010)\citenamefont {Hyllus},
  \citenamefont {G\"uhne},\ and\ \citenamefont {Smerzi}}]{Hyllus10}%
  \BibitemOpen
  \bibfield  {author} {\bibinfo {author} {\bibfnamefont {P.}~\bibnamefont
  {Hyllus}}, \bibinfo {author} {\bibfnamefont {O.}~\bibnamefont {G\"uhne}}, \
  and\ \bibinfo {author} {\bibfnamefont {A.}~\bibnamefont {Smerzi}},\ }\href
  {\doibase 10.1103/PhysRevA.82.012337} {\bibfield  {journal} {\bibinfo
  {journal} {Phys. Rev. A}\ }\textbf {\bibinfo {volume} {82}},\ \bibinfo
  {pages} {012337} (\bibinfo {year} {2010})}\BibitemShut {NoStop}%
\bibitem [{\citenamefont {T\'oth}(2012)}]{Toth12}%
  \BibitemOpen
  \bibfield  {author} {\bibinfo {author} {\bibfnamefont {G.}~\bibnamefont
  {T\'oth}},\ }\href {\doibase 10.1103/PhysRevA.85.022322} {\bibfield
  {journal} {\bibinfo  {journal} {Phys. Rev. A}\ }\textbf {\bibinfo {volume}
  {85}},\ \bibinfo {pages} {022322} (\bibinfo {year} {2012})}\BibitemShut
  {NoStop}%
\bibitem [{\citenamefont {Hayashi}\ \emph {et~al.}(2005)\citenamefont
  {Hayashi}, \citenamefont {Hashimoto},\ and\ \citenamefont
  {Horibe}}]{designs1}%
  \BibitemOpen
  \bibfield  {author} {\bibinfo {author} {\bibfnamefont {A.}~\bibnamefont
  {Hayashi}}, \bibinfo {author} {\bibfnamefont {T.}~\bibnamefont {Hashimoto}},
  \ and\ \bibinfo {author} {\bibfnamefont {M.}~\bibnamefont {Horibe}},\ }\href
  {\doibase 10.1103/PhysRevA.72.032325} {\bibfield  {journal} {\bibinfo
  {journal} {Phys. Rev. A}\ }\textbf {\bibinfo {volume} {72}},\ \bibinfo
  {pages} {032325} (\bibinfo {year} {2005})}\BibitemShut {NoStop}%
\bibitem [{\citenamefont {Dankert}\ \emph {et~al.}(2009)\citenamefont
  {Dankert}, \citenamefont {Cleve}, \citenamefont {Emerson},\ and\
  \citenamefont {Livine}}]{designs2}%
  \BibitemOpen
  \bibfield  {author} {\bibinfo {author} {\bibfnamefont {C.}~\bibnamefont
  {Dankert}}, \bibinfo {author} {\bibfnamefont {R.}~\bibnamefont {Cleve}},
  \bibinfo {author} {\bibfnamefont {J.}~\bibnamefont {Emerson}}, \ and\
  \bibinfo {author} {\bibfnamefont {E.}~\bibnamefont {Livine}},\ }\href
  {\doibase 10.1103/PhysRevA.80.012304} {\bibfield  {journal} {\bibinfo
  {journal} {Phys. Rev. A}\ }\textbf {\bibinfo {volume} {80}},\ \bibinfo
  {pages} {012304} (\bibinfo {year} {2009})}\BibitemShut {NoStop}%
\bibitem [{\citenamefont {Ko\l{}ody\ifmmode~\acute{n}\else \'{n}\fi{}ski}\ and\
  \citenamefont {Demkowicz-Dobrza\ifmmode~\acute{n}\else
  \'{n}\fi{}ski}(2010)}]{Kolodynski10}%
  \BibitemOpen
  \bibfield  {author} {\bibinfo {author} {\bibfnamefont {J.}~\bibnamefont
  {Ko\l{}ody\ifmmode~\acute{n}\else \'{n}\fi{}ski}}\ and\ \bibinfo {author}
  {\bibfnamefont {R.}~\bibnamefont {Demkowicz-Dobrza\ifmmode~\acute{n}\else
  \'{n}\fi{}ski}},\ }\href {\doibase 10.1103/PhysRevA.82.053804} {\bibfield
  {journal} {\bibinfo  {journal} {Phys. Rev. A}\ }\textbf {\bibinfo {volume}
  {82}},\ \bibinfo {pages} {053804} (\bibinfo {year} {2010})}\BibitemShut
  {NoStop}%
\bibitem [{\citenamefont {Hyllus}\ \emph {et~al.}(2012)\citenamefont {Hyllus},
  \citenamefont {Laskowski}, \citenamefont {Krischek}, \citenamefont
  {Schwemmer}, \citenamefont {Wieczorek}, \citenamefont {Weinfurter},
  \citenamefont {Pezz\'e},\ and\ \citenamefont {Smerzi}}]{Hyllus12}%
  \BibitemOpen
  \bibfield  {author} {\bibinfo {author} {\bibfnamefont {P.}~\bibnamefont
  {Hyllus}}, \bibinfo {author} {\bibfnamefont {W.}~\bibnamefont {Laskowski}},
  \bibinfo {author} {\bibfnamefont {R.}~\bibnamefont {Krischek}}, \bibinfo
  {author} {\bibfnamefont {C.}~\bibnamefont {Schwemmer}}, \bibinfo {author}
  {\bibfnamefont {W.}~\bibnamefont {Wieczorek}}, \bibinfo {author}
  {\bibfnamefont {H.}~\bibnamefont {Weinfurter}}, \bibinfo {author}
  {\bibfnamefont {L.}~\bibnamefont {Pezz\'e}}, \ and\ \bibinfo {author}
  {\bibfnamefont {A.}~\bibnamefont {Smerzi}},\ }\href {\doibase
  10.1103/PhysRevA.85.022321} {\bibfield  {journal} {\bibinfo  {journal} {Phys.
  Rev. A}\ }\textbf {\bibinfo {volume} {85}},\ \bibinfo {pages} {022321}
  (\bibinfo {year} {2012})}\BibitemShut {NoStop}%
\bibitem [{\citenamefont {Badzia\ifmmode~\mbox{\c{}}\else \c{}\fi{}g}\ \emph
  {et~al.}(2008)\citenamefont {Badzia\ifmmode~\mbox{\c{}}\else \c{}\fi{}g},
  \citenamefont {Brukner}, \citenamefont {Laskowski}, \citenamefont {Paterek},\
  and\ \citenamefont {\ifmmode~\dot{Z}\else \.{Z}\fi{}ukowski}}]{Badziag08}%
  \BibitemOpen
  \bibfield  {author} {\bibinfo {author} {\bibfnamefont {P.}~\bibnamefont
  {Badzia\ifmmode~\mbox{\c{}}\else \c{}\fi{}g}}, \bibinfo {author}
  {\bibfnamefont {C.}~\bibnamefont {Brukner}}, \bibinfo {author} {\bibfnamefont
  {W.}~\bibnamefont {Laskowski}}, \bibinfo {author} {\bibfnamefont
  {T.}~\bibnamefont {Paterek}}, \ and\ \bibinfo {author} {\bibfnamefont
  {M.}~\bibnamefont {\ifmmode~\dot{Z}\else \.{Z}\fi{}ukowski}},\ }\href
  {\doibase 10.1103/PhysRevLett.100.140403} {\bibfield  {journal} {\bibinfo
  {journal} {Phys. Rev. Lett.}\ }\textbf {\bibinfo {volume} {100}},\ \bibinfo
  {pages} {140403} (\bibinfo {year} {2008})}\BibitemShut {NoStop}%
\bibitem [{\citenamefont {Laskowski}\ \emph {et~al.}(2011)\citenamefont
  {Laskowski}, \citenamefont {Markiewicz}, \citenamefont {Paterek},\ and\
  \citenamefont {\ifmmode~\dot{Z}\else \.{Z}\fi{}ukowski}}]{Laskowski11}%
  \BibitemOpen
  \bibfield  {author} {\bibinfo {author} {\bibfnamefont {W.}~\bibnamefont
  {Laskowski}}, \bibinfo {author} {\bibfnamefont {M.}~\bibnamefont
  {Markiewicz}}, \bibinfo {author} {\bibfnamefont {T.}~\bibnamefont {Paterek}},
  \ and\ \bibinfo {author} {\bibfnamefont {M.}~\bibnamefont
  {\ifmmode~\dot{Z}\else \.{Z}\fi{}ukowski}},\ }\href {\doibase
  10.1103/PhysRevA.84.062305} {\bibfield  {journal} {\bibinfo  {journal} {Phys.
  Rev. A}\ }\textbf {\bibinfo {volume} {84}},\ \bibinfo {pages} {062305}
  (\bibinfo {year} {2011})}\BibitemShut {NoStop}%
\bibitem [{\citenamefont {Laskowski}\ \emph {et~al.}(2013)\citenamefont
  {Laskowski}, \citenamefont {Markiewicz}, \citenamefont {Paterek},\ and\
  \citenamefont {Weinar}}]{Laskowski13}%
  \BibitemOpen
  \bibfield  {author} {\bibinfo {author} {\bibfnamefont {W.}~\bibnamefont
  {Laskowski}}, \bibinfo {author} {\bibfnamefont {M.}~\bibnamefont
  {Markiewicz}}, \bibinfo {author} {\bibfnamefont {T.}~\bibnamefont {Paterek}},
  \ and\ \bibinfo {author} {\bibfnamefont {R.}~\bibnamefont {Weinar}},\ }\href
  {\doibase 10.1103/PhysRevA.88.022304} {\bibfield  {journal} {\bibinfo
  {journal} {Phys. Rev. A}\ }\textbf {\bibinfo {volume} {88}},\ \bibinfo
  {pages} {022304} (\bibinfo {year} {2013})}\BibitemShut {NoStop}%
\bibitem [{\citenamefont {Scott}(2004)}]{Scott04}%
  \BibitemOpen
  \bibfield  {author} {\bibinfo {author} {\bibfnamefont {A.~J.}\ \bibnamefont
  {Scott}},\ }\href {\doibase 10.1103/PhysRevA.69.052330} {\bibfield  {journal}
  {\bibinfo  {journal} {Phys. Rev. A}\ }\textbf {\bibinfo {volume} {69}},\
  \bibinfo {pages} {052330} (\bibinfo {year} {2004})}\BibitemShut {NoStop}%
\bibitem [{\citenamefont {Devanathan}(2002)}]{Devanathan02}%
  \BibitemOpen
  \bibfield  {author} {\bibinfo {author} {\bibfnamefont {V.}~\bibnamefont
  {Devanathan}},\ }\href@noop {} {\emph {\bibinfo {title} {Angular Momentum
  Techniques in Quantum Mechanics}}}\ (\bibinfo  {publisher} {Springer
  Netherlands},\ \bibinfo {year} {2002})\BibitemShut {NoStop}%
\bibitem [{\citenamefont {Helwig}\ and\ \citenamefont {Cui}(2013)}]{Helwig13}%
  \BibitemOpen
  \bibfield  {author} {\bibinfo {author} {\bibfnamefont {W.}~\bibnamefont
  {Helwig}}\ and\ \bibinfo {author} {\bibfnamefont {W.}~\bibnamefont {Cui}},\
  }\href {https://128.84.21.199/abs/1306.2536} {\bibfield  {journal} {\bibinfo
  {journal} {arXiv:1306.2536 [quant-ph]}\ } (\bibinfo {year}
  {2013})}\BibitemShut {NoStop}%
\bibitem [{\citenamefont {Greenberger}\ \emph {et~al.}(1989)\citenamefont
  {Greenberger}, \citenamefont {Horne},\ and\ \citenamefont {Zeilinger}}]{GHZ}%
  \BibitemOpen
  \bibfield  {author} {\bibinfo {author} {\bibfnamefont {D.~M.}\ \bibnamefont
  {Greenberger}}, \bibinfo {author} {\bibfnamefont {M.~A.}\ \bibnamefont
  {Horne}}, \ and\ \bibinfo {author} {\bibfnamefont {A.}~\bibnamefont
  {Zeilinger}},\ }\href {https://arxiv.org/abs/0712.0921} {\emph {\bibinfo
  {title} {Bells Theorem, Quantum Theory and Conceptions of the Universe}}}\
  (\bibinfo  {publisher} {Kluwer, Dordrecht},\ \bibinfo {year}
  {1989})\BibitemShut {NoStop}%
\bibitem [{\citenamefont {Dicke}(1954)}]{Dicke54}%
  \BibitemOpen
  \bibfield  {author} {\bibinfo {author} {\bibfnamefont {R.~H.}\ \bibnamefont
  {Dicke}},\ }\href {\doibase http://dx.doi.org/10.1103/PhysRev.93.99}
  {\bibfield  {journal} {\bibinfo  {journal} {Phys. Rev.}\ }\textbf {\bibinfo
  {volume} {93}},\ \bibinfo {pages} {99} (\bibinfo {year} {1954})}\BibitemShut
  {NoStop}%
\bibitem [{\citenamefont {Borras}\ \emph {et~al.}(2007)\citenamefont {Borras},
  \citenamefont {Plastino}, \citenamefont {Batle}, \citenamefont {Zander},
  \citenamefont {Casas},\ and\ \citenamefont {Plastino}}]{Borras07}%
  \BibitemOpen
  \bibfield  {author} {\bibinfo {author} {\bibfnamefont {A.}~\bibnamefont
  {Borras}}, \bibinfo {author} {\bibfnamefont {A.~R.}\ \bibnamefont
  {Plastino}}, \bibinfo {author} {\bibfnamefont {J.}~\bibnamefont {Batle}},
  \bibinfo {author} {\bibfnamefont {C.}~\bibnamefont {Zander}}, \bibinfo
  {author} {\bibfnamefont {M.}~\bibnamefont {Casas}}, \ and\ \bibinfo {author}
  {\bibfnamefont {A.}~\bibnamefont {Plastino}},\ }\href
  {http://stacks.iop.org/1751-8121/40/i=44/a=018} {\bibfield  {journal}
  {\bibinfo  {journal} {J. Phys. A: Math. Theor.}\ }\textbf {\bibinfo {volume}
  {40}},\ \bibinfo {pages} {13407} (\bibinfo {year} {2007})}\BibitemShut
  {NoStop}%
\bibitem [{\citenamefont {Goyeneche}\ \emph {et~al.}(2015)\citenamefont
  {Goyeneche}, \citenamefont {Alsina}, \citenamefont {Latorre}, \citenamefont
  {Riera},\ and\ \citenamefont {\.Zyczkowski}}]{AME}%
  \BibitemOpen
  \bibfield  {author} {\bibinfo {author} {\bibfnamefont {D.}~\bibnamefont
  {Goyeneche}}, \bibinfo {author} {\bibfnamefont {D.}~\bibnamefont {Alsina}},
  \bibinfo {author} {\bibfnamefont {J.~I.}\ \bibnamefont {Latorre}}, \bibinfo
  {author} {\bibfnamefont {A.}~\bibnamefont {Riera}}, \ and\ \bibinfo {author}
  {\bibfnamefont {K.}~\bibnamefont {\.Zyczkowski}},\ }\href {\doibase
  10.1103/PhysRevA.92.032316} {\bibfield  {journal} {\bibinfo  {journal} {Phys.
  Rev. A}\ }\textbf {\bibinfo {volume} {92}},\ \bibinfo {pages} {032316}
  (\bibinfo {year} {2015})}\BibitemShut {NoStop}%
\bibitem [{\citenamefont {Krischek}\ \emph {et~al.}(2011)\citenamefont
  {Krischek}, \citenamefont {Schwemmer}, \citenamefont {Wieczorek},
  \citenamefont {Weinfurter}, \citenamefont {Hyllus}, \citenamefont {Pezz\'e},\
  and\ \citenamefont {Smerzi}}]{Krischek11}%
  \BibitemOpen
  \bibfield  {author} {\bibinfo {author} {\bibfnamefont {R.}~\bibnamefont
  {Krischek}}, \bibinfo {author} {\bibfnamefont {C.}~\bibnamefont {Schwemmer}},
  \bibinfo {author} {\bibfnamefont {W.}~\bibnamefont {Wieczorek}}, \bibinfo
  {author} {\bibfnamefont {H.}~\bibnamefont {Weinfurter}}, \bibinfo {author}
  {\bibfnamefont {P.}~\bibnamefont {Hyllus}}, \bibinfo {author} {\bibfnamefont
  {L.}~\bibnamefont {Pezz\'e}}, \ and\ \bibinfo {author} {\bibfnamefont
  {A.}~\bibnamefont {Smerzi}},\ }\href {\doibase
  10.1103/PhysRevLett.107.080504} {\bibfield  {journal} {\bibinfo  {journal}
  {Phys. Rev. Lett.}\ }\textbf {\bibinfo {volume} {107}},\ \bibinfo {pages}
  {080504} (\bibinfo {year} {2011})}\BibitemShut {NoStop}%
\end{thebibliography}

\end{document}